\documentclass[useAMS,usenatbib ]{mn2e}
\usepackage{graphicx}
\usepackage{amsmath}
\usepackage{pdfpages}
\usepackage{longtable}

\usepackage{color}
\definecolor{darkgreen}{rgb}{0,0.6,0}
\definecolor{gray}{rgb}{0.5,0.5,0.5}
\definecolor{mauve}{rgb}{0.58,0,0.82}
\usepackage[T1]{fontenc}
\usepackage{aecompl}
\usepackage{hyperref}
\usepackage{epstopdf}
\hypersetup{breaklinks,colorlinks,citecolor=blue}

\newif\ifdraft
\drafttrue

\ifdraft
  \newcommand{\before}[1]{\textcolor{darkgreen}{[{\bf BEFORE}: #1]}}

\else
  \newcommand{\before}[1]{}
  
\fi

\title[NIR light curves of Type Ia supernovae]{Near-infrared light curves of Type Ia supernovae: Studying properties of the second maximum}
\author[S. Dhawan \emph{et al.}]{S. Dhawan$^{1,2,3}$\thanks{sdhawan@eso.org}, 
B. Leibundgut$^{1,2}$, J. Spyromilio$^1$, K. Maguire$^1$\\
$^1$European Southern Observatory, Karl-Schwarzschild-Strasse, 2, D-85748 Garching bei M\"unchen, Germany\\
$^2$Excellence Cluster Universe, Technische Universit\"at M\"unchen,
Boltzmannstrasse 2, D-85748, Garching, Germany\\
$^3$Physik Department, Technische Universit\"at M\"unchen, James-Franck-Strasse 1, D-85748 Garching bei M\"unchen\\
}

\date{Accepted ---. Received ---; in original form ---}
\pagerange{\pageref{firstpage}--\pageref{lastpage}} 
\pubyear{2014}

\begin{document}
\maketitle
\label{firstpage}
\begin{abstract}
Type Ia supernovae have been proposed to be much better distance indicators at near-infrared  compared to optical wavelengths -- the effect of dust extinction is expected to be lower and it has been shown that SNe Ia behave more like `standard candles' at NIR wavelengths. To better understand the physical processes behind this increased uniformity, we have studied the $Y$, $J$ and $H$-filter light curves of 91 SNe Ia from the literature. We show that the phases and luminosities of the first maximum in the NIR  light curves are extremely uniform for our sample. 
The phase of the second maximum, the late-phase NIR  luminosity and the optical light curve shape are found to be strongly correlated, in particular more luminous SNe Ia reach the second maximum in the NIR  filters at a later phase compared to fainter objects. We also find a strong correlation between the phase of the second maximum and the epoch at which the SN enters the Lira law phase in its optical colour curve (epochs $\sim$ 15 to 30 days after $B$ band maximum). The decline rate after the second maximum is very uniform in all NIR  filters.  We suggest that these observational parameters are linked to the nickel and iron mass in the
explosion, providing evidence that the amount of nickel synthesised in
the explosion is the dominating factor shaping the optical and NIR 
appearance of SNe Ia. 

\end{abstract}


\section{Introduction}
\label{sec-intro}

The uniformity of Type Ia supernovae (SNe Ia) has led to their use, after calibration, as
distance indicators \citep[reviewed  
in:][]{Goobar2011} and they provided the first evidence for the accelerated
expansion of the Universe \citep{Riess1998,Perlmutter1999}. 

Observations of large SN Ia samples
show that the peak luminosity in the optical is not uniform but can be normalised following a variety of calibration techniques, most notably the correlation between light curve shape and peak luminosity, and between light curve colour and peak luminosity
\citep[e.g.][]{Phillips1993, Riess1996, Guy2005, Guy2007, Guy2010,
Jha2007}. 
The variation in bolometric luminosity for the objects 
\citep{Contardo2000} implies variations in the physical
parameters of the explosion, in particular the synthesised Ni mass
and the total ejected mass \citep{Stritzinger2006, Scalzo2014}. 

At near-infrared (NIR) wavelengths  (900nm $>\lambda >$ 2000nm), SNe Ia have a very uniform brightness
distribution without any prior normalisation \citep{Elias1981, Meikle2000, K04a, K07}. 
The scatter in the peak luminosity in these studies is $\sim$0.2 mag, which when combined
with the lower sensitivity of the NIR to extinction by dust, has sparked
 interest in the use of this wavelength region. Following large observational campaigns, statistically significant
samples of SN~Ia light curves have been made public \citep{WV08,
Contreras2010, Stritzinger2011, BN12}, and have been used to construct
the first rest-frame NIR Hubble diagrams
\citep{Nobili2005,Freedman2009, Kattner2012, Weyant2014}. 

Dust extinction in the NIR is significantly reduced compared
to the optical leading to smaller corrections and uncertainties. In addition to the photometric calibration systematics \citep[see][]{Conley2011} the dust extinction for SNe Ia is one of the major sources of systematics  in SN Ia cosmology measurements \citep[e.g.][]{Peacock2006, Goobar2011}. In particular, extinction law measurements for galaxies remains uncertain  \citep[see
discussions in][]{Phillips2013, Scolnic2014}.  The recent SN\,2014J is a case
in point with derived dust properties very different from local
interstellar dust \citep{Amanullah2014, Foley2014}.  Strongly reddened SNe Ia may also exhibit variations in their light curve shapes \citep{Leibundgut1988, Amanullah2014}. 

The light-curve morphology in the NIR is markedly different from
that in the optical with a pronounced second maximum in $IYJHK$
filters for `normal' SNe Ia  \citep{Elias1981, Elias1985, Leibundgut1988,
Leibundgut2000, Meikle2000, WV08, Folatelli2010}. 
The formation of the NIR spectrum in SNe Ia is highly sensitive to the opacity variations \citep{Spyr1994, Wheeler1998} as the spectrum is  dominated by line blanketing opacity making the evolution of the NIR light a sensitive probe of the structure of the ejecta. 
\cite{Kasen2006} in a detailed study suggested that the second maximum is a result of a 
decrease in opacity due to the ionisation change of Fe group elements from doubly to
singly ionised atoms, which preferentially radiate the energy at near-IR
wavelengths. A direct prediction of \citet{Kasen2006} is that a larger iron mass  
leads to a later NIR second maximum.

Studies of the $i$ 
band light curve find a relation between the phase
of the second maximum and the optical light curve shape 
\citep[e.g. $\Delta m_{15}(B)$, ][]{Folatelli2010, Hamuy1996}. The strength
of the second maximum in $i$ 
does not show such a correlation. 

In this paper, we investigate the properties of SN Ia NIR light curves ($YJH$) and
establish correlations with other observational characteristics. Connections to possible physical properties in the
explosions are explored. The structure of this paper is as follows: after a
presentation of the input data in Section \ref{sec:data}, we analyse the
NIR light curve properties (Section~\ref{sec-LC}) along with a
description of NIR colours. Correlations with optical light curve
parameters and their interpretations  are given in 
\S\ref{sec-corr} followed by a discussion in \S\ref{sec-disc}. The conclusions are presented in Section~\ref{sec-conc}.

\section{Data} 
\label{sec:data}

\begin{table}
\tiny
\begin{minipage}{75mm}
\begin{center}
\caption{SN sample \label{tab:sne_ref}}
\begin{tabular}{l r@{\ } c @{}r r r r r r l}
\hline
SN name & \multicolumn{3}{c}{Phase range} & $N_i$ & $N_Y$ & $N_J$ & $N_H$ & $N_K$ & 
Reference\footnotetext{References: M00 - \citet{Meikle2000}; Ph06 -
\citet{Phillips2006}; K04a - \citet{K04a}; V03 - \citet{Valentini2003};
K04b - \citet{K04b}; P08 - \citet{Pignata2008}; C13 -
\citet{Cartier2013}; ER06 - \citet{ER06}; K09 - \citet{Krisciunas2009};
P07 - \citet{Pastorello2007}; CSP - Carnegie Supernova Project
\citet{Contreras2010, Stritzinger2011}; M12 - \citet{Matheson2012}; F14
- \citet{Foley2014} } \\ 
 & \multicolumn{3}{c}{(days)} & & & & & & \\
\hline
SN1980N  & $  5.7$ & $\cdots$ & $ 99.7$ & \ldots & \ldots & 11 & 13 & \ldots & M00 \\
SN1981B	 & $  2.7$ & $\cdots$ & $120.5$ & \ldots & \ldots & 17 & 17 & \ldots & M00 \\
SN1986G	 & $ -6.1$ & $\cdots$ & $101.0$ & \ldots & \ldots & 28 & 29 & \ldots & M00 \\
SN1998bu & $ -8.5$ & $\cdots$ & $ 31.5$ & \ldots & \ldots & 23 & 23 & \ldots & M00 \\
SN1999ac & $-13.1$ & $\cdots$ & $ 63.8$ & \ldots & \ldots & 30 & 30 & \ldots & Ph06 \\
SN1999ee & $ -7.5$ & $\cdots$ & $ 27.5$ & \ldots &     17 & 18 & 20 & \ldots & K04a \\
SN1999ek & $ -8.0$ & $\cdots$ & $ 23.0$ & \ldots & \ldots & 14 & 15 & \ldots & K04a \\
SN2000E	 & $ -7.7$ & $\cdots$ & $126.6$ & \ldots & \ldots & 18 & 18 & \ldots & V03 \\
SN2000bh & $ -5.5$ & $\cdots$ & $ 39.0$ & \ldots &      6 & 21 & 22 & \ldots & K04a \\
SN2001ba & $ -6.0$ & $\cdots$ & $ 34.9$ & \ldots & \ldots & 14 & 15 & \ldots & K04a \\
SN2001bt & $ -1.6$ & $\cdots$ & $ 65.3$ & \ldots & \ldots & 21 & 21 & \ldots & K04a \\
SN2001cn & $  4.2$ & $\cdots$ & $ 59.1$ & \ldots & \ldots & 19 & 19 & \ldots & K04b \\
SN2001cz & $ -2.4$ & $\cdots$ & $ 47.6$ & \ldots & \ldots & 12 & 12 & \ldots & K04b \\
SN2001el & $-10.6$ & $\cdots$ & $ 64.3$ & \ldots & \ldots & 33 & 32 & \ldots & K03 \\
SN2002bo & $-11.0$ & $\cdots$ & $ 44.0$ & \ldots & \ldots & 17 & 17 &	17   & K04b, ESC\\
SN2002dj & $-11.0$ & $\cdots$ & $ 67.0$ & \ldots & \ldots & 21 & 21 & \ldots & P08, ESC \\
SN2002fk & $-12.2$ & $\cdots$ & $102.1$ & \ldots & \ldots & 24 & 23 &	23   & C13 \\ 
SN2003cg & $ -6.4$ & $\cdots$ & $413.5$ & \ldots & \ldots & 13 & 13 & \ldots & ER06, ESC \\
SN2003hv & $  1.2$ & $\cdots$ & $ 62.0$ & \ldots &     16 & 16 & 16 & \ldots & L09, ESC \\
SN2004ef & $ -8.7$ & $\cdots$ & $ 65.2$ & 46     &      4 &  3 &  4 &      3 & CSP \\
SN2004eo & $-12.0$ & $\cdots$ & $ 63.0$ & 39     &      8 &  9 &  9 &      8 & CSP, P07, ESC \\
SN2004ey & $ -8.9$ & $\cdots$ & $ 48.1$ & 32     &      7 &  9 &  9 &      8 & CSP \\
SN2004gs & $ -3.6$ & $\cdots$ & $ 99.1$ & 50     &     12 & 11 & 10 & \ldots & CSP \\
SN2004gu & $ -0.4$ & $\cdots$ & $ 48.6$ & 27     &      8 &  7 &  7 & \ldots & CSP \\
SN2005A	 & $ -3.5$ & $\cdots$ & $ 30.4$ & 35     &     10 & 10 & 10 & \ldots & CSP \\
SN2005M	 & $ -8.0$ & $\cdots$ & $ 74.7$ & 56     &     17 & 17 & 14 &     12 & CSP \\
SN2005ag & $ -1.7$ & $\cdots$ & $ 66.3$ & 44     &      9 &  9 &  9 & \ldots & CSP \\
SN2005al & $ -1.0$ & $\cdots$ & $ 82.0$ & 35     &      7 &  8 &  8 &      7 & CSP \\
SN2005am & $ -4.6$ & $\cdots$ & $ 75.3$ & 36     &      6 &  6 &  6 &      6 & CSP \\
SN2005el & $ -7.3$ & $\cdots$ & $ 83.6$ & 25     &     21 & 22 & 15 &      3 & CSP \\
SN2005eq & $ -3.6$ & $\cdots$ & $ 97.3$ & 27     &     15 & 15 & 10 &      1 & CSP \\
SN2005hc & $ -4.5$ & $\cdots$ & $ 84.4$ & 22     &     13 & 11 &  9 & \ldots & CSP \\
SN2005hj & $ -2.3$ & $\cdots$ & $ 87.6$ & 16     &     11 & 12 & 10 &      1 & CSP \\
SN2005iq & $ -5.3$ & $\cdots$ & $ 67.7$ & 19     &     11 & 11 & 11 &      1 & CSP \\
SN2005kc & $-10.4$ & $\cdots$ & $ 30.6$ & 13     &      9 &  9 &  8 & \ldots & CSP \\
SN2005ki & $ -9.9$ & $\cdots$ & $155.8$ & 47     &     12 & 11 & 10 & \ldots & CSP \\
SN2005na & $ -1.8$ & $\cdots$ & $ 94.0$ & 27     &     14 & 11 & 12 & \ldots & CSP \\
SN2006D	 & $ -6.0$ & $\cdots$ & $112.8$ & 42     &     17 & 16 & 16 &      6 & CSP \\
SN2006X	 & $-11.0$ & $\cdots$ & $119.8$ & 39     &     32 & 33 & 32 &      9 & CSP, ESC \\
SN2006ax & $-11.7$ & $\cdots$ & $ 71.2$ & 26     &     19 & 18 & 16 &      3 & CSP \\
SN2006bh & $ -4.9$ & $\cdots$ & $ 60.0$ & 24     &     12 & 11 & 10 & \ldots & CSP \\
SN2006br & $  5.8$ & $\cdots$ & $ 37.7$ &  9     &      5 &  5 &  5 & \ldots & CSP \\
SN2006ej & $  4.4$ & $\cdots$ & $ 69.4$ & 13     &      3 &  3 &  3 & \ldots & CSP \\
SN2006eq & $  0.2$ & $\cdots$ & $ 43.0$ & 18     &     10 &  7 &  8 & \ldots & CSP \\
SN2006et & $ -7.2$ & $\cdots$ & $105.7$ & 23     &     18 & 12 & 13 & \ldots & CSP \\
SN2006ev & $  4.2$ & $\cdots$ & $ 48.1$ & 12     &     10 &  8 &  8 & \ldots & CSP \\
SN2006gj & $ -1.8$ & $\cdots$ & $ 96.0$ & 19     &     13 & 10 &  4 & \ldots & CSP \\
SN2006gt & $ -2.1$ & $\cdots$ & $ 62.8$ & 13     &     10 &  8 &  6 & \ldots & CSP \\
SN2006hb & $  7.2$ & $\cdots$ & $136.9$ & 25     &     10 & 10 &  9 & \ldots & CSP \\
SN2006hx & $ -9.3$ & $\cdots$ & $ 30.7$ &  8     &      7 &  6 &  5 & \ldots & CSP \\
SN2006is & $  6.4$ & $\cdots$ & $118.2$ & 24     &      8 &  8 &  7 & \ldots & CSP \\
SN2006kf & $ -4.8$ & $\cdots$ & $ 86.0$ & 20     &     17 & 14 & 11 &      3 & CSP \\
SN2006lu & $  5.1$ & $\cdots$ & $ 90.1$ & 21     &      6 &  4 &  3 & \ldots & CSP \\
SN2006ob & $ -2.6$ & $\cdots$ & $ 58.3$ & 13     &     12 &  9 &  5 & \ldots & CSP \\
SN2006os & $  2.4$ & $\cdots$ & $ 57.3$ & 14     &     10 &  6 &  5 & \ldots & CSP \\
SN2007A	 & $ -4.7$ & $\cdots$ & $ 18.3$ &  9     &      9 &  5 &  3 & \ldots & CSP \\
SN2007S	 & $-10.5$ & $\cdots$ & $102.2$ & 19     &     12 & 17 & 18 &      7 & CSP \\
SN2007af & $ -8.8$ & $\cdots$ & $ 82.1$ & 28     &     26 & 25 & 24 &      5 & CSP \\
SN2007ai & $ -0.4$ & $\cdots$ & $ 86.4$ & 17     &      7 &  7 &  6 &      3 & CSP \\
SN2007as & $ -0.0$ & $\cdots$ & $ 77.7$ & 19     &     11 & 10 & 10 & \ldots & CSP \\
SN2007bc & $ -3.0$ & $\cdots$ & $ 58.8$ & 10     &     11 &  8 &  6 & \ldots & CSP \\
SN2007bd & $ -9.5$ & $\cdots$ & $215.8$ & 14     &     12 &  9 &  7 & \ldots & CSP \\
SN2007bm & $ -0.2$ & $\cdots$ & $ 34.7$ & 10     &     10 &  9 &  7 & \ldots & CSP \\
SN2007ca & $ -8.8$ & $\cdots$ & $ 32.1$ & 12     &     10 &  8 &  7 &      2 & CSP \\
SN2007if & $ 14.4$ & $\cdots$ & $100.2$ & 16     &      8 &  7 &  5 & \ldots & CSP \\
SN2007jg & $ -3.1$ & $\cdots$ & $ 58.8$ & 18     &      8 &  6 &  5 & \ldots & CSP \\
SN2007le & $-11.5$ & $\cdots$ & $ 71.2$ & 26     &     17 & 17 & 16 & \ldots & CSP \\
SN2007nq & $ -1.4$ & $\cdots$ & $ 77.5$ & 25     &     19 & 10 &  4 & \ldots & CSP \\
SN2007on & $ -8.6$ & $\cdots$ & $ 88.2$ & 38     &     29 & 28 & 25 &      7 & CSP \\
SN2008C	 & $  2.9$ & $\cdots$ & $ 85.6$ & 19     &     15 & 13 & 18 &      1 & CSP \\
SN2008R	 & $ -1.3$ & $\cdots$ & $ 32.6$ & 12     &      8 &  7 &  5 & \ldots & CSP \\
SN2008bc & $ -9.3$ & $\cdots$ & $ 96.6$ & 32     &      7 & 12 & 11 & \ldots & CSP \\
SN2008bq & $ -0.8$ & $\cdots$ & $ 43.1$ & 16     &      4 &  4 &  4 & \ldots & CSP \\
SN2008fp & $ -6.0$ & $\cdots$ & $ 89.9$ & 28     &     22 & 20 & 20 &      7 & CSP \\
SN2008gp & $-10.5$ & $\cdots$ & $ 34.4$ & 19     &     10 & 11 &  9 & \ldots & CSP \\
SN2008hv & $-10.5$ & $\cdots$ & $ 78.3$ & 25     &     18 & 16 & 16 & \ldots & CSP \\
SN2008ia & $ -3.5$ & $\cdots$ & $ 43.5$ & 15     &     16 & 15 & 14 & \ldots & CSP \\
PTF09dlc & $-5.56$ & $\cdots$ & $ 19.5$ & \ldots & \ldots &  4 &  4 & \ldots & BN12 \\
PTF10hdv & $ -3.9$ & $\cdots$ & $ 10.3$ & \ldots & \ldots &  4 &  4 & \ldots & BN12 \\ 
PTF10hmv & $-10.1$ & $\cdots$ & $  7.9$ & \ldots & \ldots & \ldots & 5 & \ldots & BN12 \\ 
PTF10mwb & $-12.3$ & $\cdots$ & $  5.7$ & \ldots & \ldots &  5 &  5 & \ldots & BN12 \\ 
PTF10ndc & $ -3.9$ & $\cdots$ & $  5.1$ & \ldots & \ldots &  4 &  4 & \ldots & BN12 \\ 
PTF10nlg & $ -5.1$ & $\cdots$ & $  3.9$ & \ldots & \ldots &  3 &  5 & \ldots & BN12 \\ 
PTF10qyx & $ -3.5$ & $\cdots$ & $  7.5$ & \ldots & \ldots &  4 &  4 & \ldots & BN12 \\ 
PTF10tce & $ -6.6$ & $\cdots$ & $  6.5$ & \ldots & \ldots &  4 &  4 & \ldots & BN12 \\ 
PTF10ufj & $ -9.0$ & $\cdots$ & $  4.1$ & \ldots & \ldots &  4 &  4 & \ldots & BN12 \\ 
PTF10wnm & $ -7.1$ & $\cdots$ & $  4.8$ & \ldots & \ldots &  4 &  4 & \ldots & BN12 \\ 
PTF10wof & $ -3.9$ & $\cdots$ & $  9.1$ & \ldots & \ldots &  4 &  4 & \ldots & BN12 \\ 
PTF10xyt & $ -4.6$ & $\cdots$ & $  7.3$ & \ldots & \ldots &  2 &  5 & \ldots & BN12 \\ 
SN2011fe & $-16.0$ & $\cdots$ & $ 45.9$ & \ldots & \ldots & 32 & 35 &     32 & M12 \\
SN2014J	 & $-10.0$ & $\cdots$ & $ 72.4$ & \ldots & \ldots & 24 & 24 &     24 & F14 \\	
\hline
\end{tabular}
\end{center}
\end{minipage}
\end{table}

We investigate a large sample of nearby objects with well-sampled
optical and NIR data (Table~\ref{tab:sne_ref}). The main data source of
NIR SN Ia photometry is the Carnegie
SN Project 
\citep[CSP;][]{Contreras2010, Burns2011,
Stritzinger2011, Phillips2012, Burns2014}.  The low-redshift CSP provides a sample of SNe Ia with
optical and NIR light curves in a homogeneous and well-defined
photometric system (in Vega magnitude system) and thus forms an ideal basis for the evaluation of light curve properties. CSP relies primarily on SN discoveries from the Lick
Observatory SN Search \citep[LOSS;][]{Leaman2011}.  The CSP has
published light curves on a total of 82 SNe Ia of which 70 have
photometry in $YJHK$ bands. 

From the CSP NIR dataset, we removed spectroscopically peculiar objects
such as SN2006bt and SN2006ot. We also rejected SNe Ia with spectra
similar to the peculiar SN~1991bg 
\citep{Filippenko1992, Leibundgut1993,
Mazzali1997} and objects that do not exhibit a second maximum
(SNe~2005bl, 2005ke, 2005ku, 2006bd, 2006mr, 2007N, 2007ax, 2007ba,
2009F). 

We have included in our sample near-IR SN\,Ia photometry from \cite{Meikle2000} and several
SNe Ia observed by the European SN Consortium
\citep[ESC;][]{Benetti2004, Pignata2008, ER06, Pastorello2007,
Krisciunas2009}. 
Twelve SNe Ia have
been discussed by \citet{BN12} and have data only near the first
maximum. We also included NIR photometry from two recent nearby
explosions, SN2011fe \citep{Matheson2012} and SN2014J \citep{Foley2014}. 

The  91 objects used in this work are listed in
Table~\ref{tab:sne_ref}, where the phase range of observations (first and
last observation), total number of observations in each filter and the
reference for each data set are tabulated. The sample is dominated by
SNe Ia from the CSP and we show the
results separately for the CSP and non-CSP objects. It is worth noting that there are
15 SNe Ia with observed NIR light curves beyond 100 days.

As can be seen in Table~\ref{tab:sne_ref} and displayed in
Fig.~\ref{fig:lc1}, the $K$-band light curves are sparsely sampled and
not enough objects are available for detailed analysis. Therefore, we 
exclude the \textit{K}-band light curves from further analysis.

\section{NIR Light Curve Morphology}
\label{sec-LC}

In parameterising the NIR light curves 
we follow the nomenclature introduced by \citet{Biscardi2012}. The first
maximum in filter $X$, $M_1(X)$ is reached at a phase, $t_1(X)$ relative
to the phase of the $B$ maximum ($t^{max}_{B}$=0 d). The light curves dip to a
minimum $M_0(X)$ at $t_0(X)$ before reaching a second maximum,
$M_2(X)$ at time, $t_2(X)$. These three prominent features are shown in Fig.~\ref{fig:lc1}. The light
curves are plotted without normalisation for phase or corrections for possible
differences in photometric systems or absorption.

\subsection{Light curve fitting}
\label{ssec-lcfit}

We fit the optical light curves using the programme SNooPy
\citep{Burns2011} to determine the peak of the $B$-band light curve
through cubic-spline fitting and determine the phase, $t^{max}_{B}$ used throughout this work and
the peak brightness $m^{max}_{B}$. SNooPY also determines the $\Delta
m_{15}$ parameter\footnote{The $\Delta m_{15}$ is calculated by SNooPy from all available filters. It is linearly related
 to $\Delta m_{15}(B)$  \citep[see][]{Burns2011}}
 commonly used to characterise the SN light
curve shape and provides an estimate of the extinction in the
 host galaxy. 
The SNooPy light curve fit parameters and distance moduli for the
SNe\,Ia in our sample are given in Table~\ref{tab:snpy}. We used the published values of the distance modulus, $\mu$, for non-CSP objects, 
(references in Table~\ref{tab:snpy}). The distance moduli for CSP
SNe\,Ia not in the Hubble flow are taken from
\citet{Contreras2010} and \citet{Stritzinger2011}, and  
the individual references 
are listed in
Table ~\ref{tab:snpy}. The distance moduli for the rest of the
SNe are based on the host galaxy redshift from the NASA/IPAC Extragalactic Database 
adopting a
Hubble constant of $H_0$=70~km~s$^{-1}$~Mpc$^{-1}$. 
\begin{figure*}
\centering
\includegraphics[width=.98\textwidth, height=0.65\textheight]{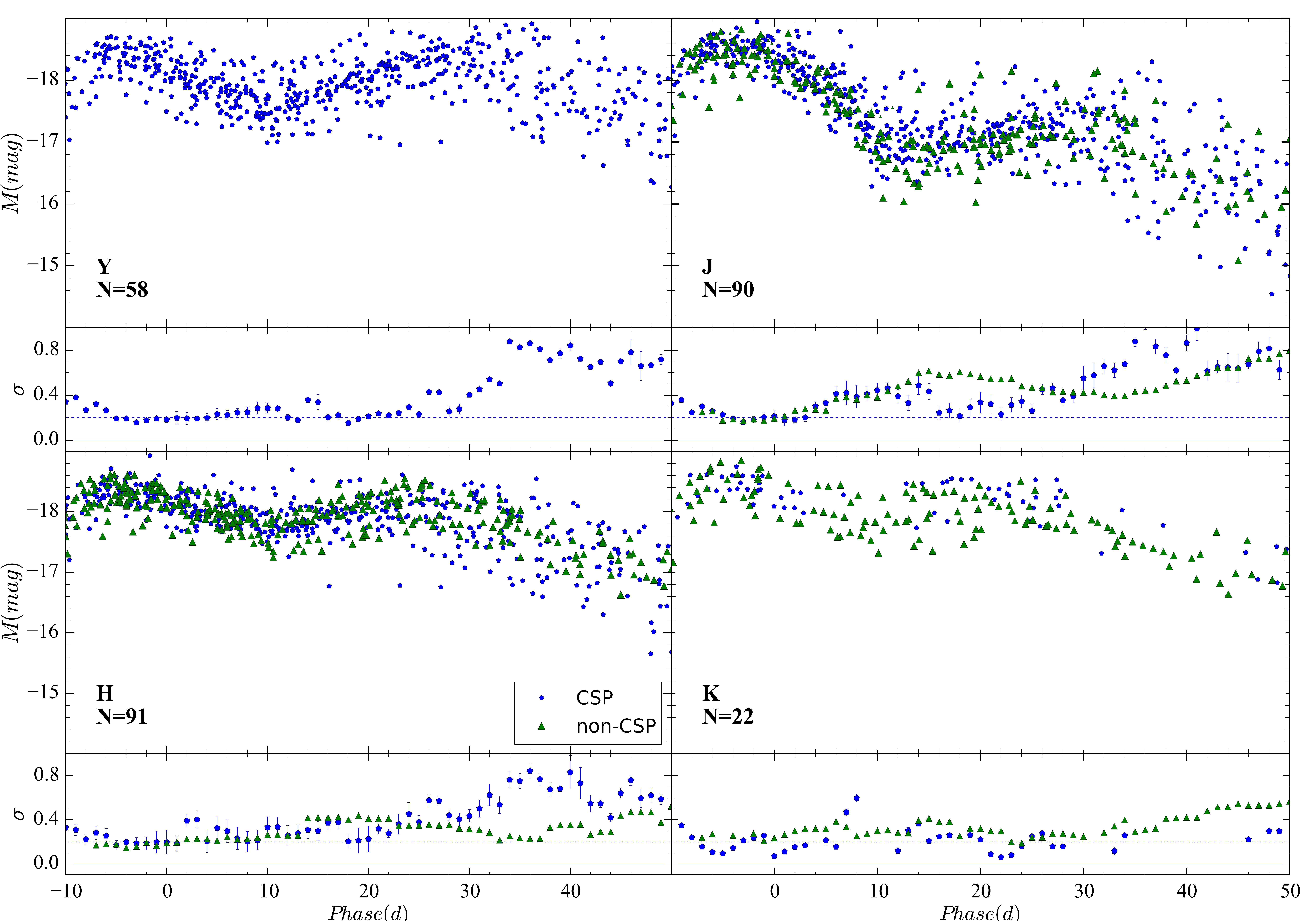}
\caption{
The $Y$, $J$, $H$ and $K$ absolute magnitude light curves of SNe\,Ia. The CSP sample is
shown as blue pentagons. The green triangles indicate the non-CSP sample (see Table~\ref{tab:sne_ref} for the
references). The root-mean-square (RMS) scatter for each day is shown in the lower panels.
For guidance a line at a scatter of 0.2 magnitudes is drawn in each
panel. The luminosities are based on the literature distances (cf.
Table~\ref{tab:snpy}) and no correction has been applied. The steady increase
of the scatter after the first maximum is evident.
}
\label{fig:lc1}
\end{figure*}

\begin{table}
\tiny
\begin{minipage}{70mm}
\begin{center}
\caption{
Timing of $B_{max}$, $\Delta m_{15}$, distance modulus for the SN in our
sample. References for the published distance moduli are provided in the
footnote.
}
\begin{tabular}{l l l l l l l }
\hline
SN Name & \multicolumn{1}{c}{$t(B_{max})$} & err & 
\multicolumn{1}{c}{$\Delta m_{15}$} & err & \multicolumn{1}{c}{$\mu$} & 
err\footnotetext{References: SN1980N, SN1981B, SN1986G, SN1998bu 
\citep{Meikle2000}; SN1999ac, SN1999ee, SN2002dj \citep{Willick1997}; 
SN2000E, SN2002bo \citep{Tully1988}; SN2001el \citep{Ajhar2001};  
SN2003hv, SN2007on \citep{Tonry2001}; 2006X \citep{Freedman2001} 
(note: we add the correction to $\mu$'s from \citet{Tonry2001} and 
\citet{Ajhar2001} using the values provided in \citet{Jensen2003}) 
} \\
 & \multicolumn{1}{c}{(MJD)} & & \multicolumn{1}{c}{(mag)} & & \multicolumn{1}{c}{(mag)} & \\
\hline
SN1980N		&	44585.8	&	0.5	&	1.28	&	0.04	&	31.59	&	0.10	\\ 
SN1981B		&	44672.0	&	0.2	&	1.10	&	0.04	&	30.95	&	0.07	\\ 
SN1986G		& 	46561.0 & 	0.1	& 	1.76 	&	0.10	&	28.01	&  	0.12  	\\
SN1998bu	&	50953.3	&	0.5	&	1.01	&	0.02	&	30.20	&	0.10	\\ 
SN1999ac	&	51251.0	&	0.5	&	1.34	&	0.02	&	33.50	&	0.30	\\ 
SN1999ee	&	51469.1	&	0.5	&	1.09	&	0.02	&	33.20	&	0.20	\\ 
SN1999ek	&	51481.8	&	0.1	&	1.17	&	0.03	&	34.40	&	0.30	\\ 
SN2000E		&	51557.0	&	0.5	&	0.99	&	0.02	&	31.90	&	0.40	\\ 
SN2000bh	&	51635.2	&	0.5	&	1.16	&	0.01	&	34.60	&	0.30	\\ 
SN2001ba	&	52034.5	&	0.5	&	0.97	&	0.05	&	35.40	&	0.50	\\ 
SN2001bt	&	52063.4	&	0.5	&	1.18	&	0.02	&	34.10	&	0.40	\\ 
SN2001cn	&	52071.6	&	0.2	&	1.15	&	0.02	&	34.10	&	0.30	\\ 
SN2001cz	&	52103.9	&	0.1	&	1.05	&	0.07	&	33.50   &	0.10	\\ 
SN2001el	&	52182.5	&	0.5	&	1.13	&	0.04	&	31.30	&	0.20 	\\ 
SN2002bo	&	52356.5	&	0.2	&	1.12	&	0.02	&	31.80	&	0.20	\\ 
SN2002dj	&	52450.0	&	0.7	&	1.08	&	0.02	&	32.90	&	0.30	\\ 
SN2002fk	&	52547.9	&	0.3	&	1.02	&	0.04	&	32.59	&	0.15	\\ 
SN2003cg	&	52729.4	&	0.5	&	1.12	&	0.04	&	31.28	&	0.20	\\ 
SN2003hv	&	52891.2	&	0.3	&	1.09	&	0.02	&	31.40	&	0.30	\\ 
SN2004ef	&	53264.4	&	0.1	&	1.45	&	0.01	&	35.57	&	0.07	\\ 
SN2004eo	&	53278.4	&	0.1	&	1.32	&	0.01	&	34.03	&	0.10	\\ 
SN2004ey	&	53304.3	&	0.1	&	1.02	&	0.01	&	34.01	&	0.12	\\ 
SN2004gs	&	53356.2	&	0.1	&	1.53	&	0.01	&	35.40	&	0.08	\\ 
SN2004gu	&	53362.2	&	0.2	&	0.80	&	0.01	&	36.59	&	0.04	\\ 
SN2005A		&	53379.7	&	0.2	&	1.08	&	0.02	&	34.51	&	0.11	\\ 
SN2005M		&	53405.4	&	0.1	&	0.80	&	0.04	&	35.01	&	0.09	\\ 
SN2005ag	&	53413.7	&	0.2	&	0.87	&	0.01	&	37.80	&	0.03	\\ 
SN2005al	&	53430.5	&	0.1	&	1.30	&	0.01	&	33.79	&	0.15	\\ 
SN2005am	&	53436.9	&	0.1	&	1.48	&	0.01	&	32.85	&	0.20	\\ 
SN2005el	&	53647.0	&	0.1	&	1.40	&	0.01	&	34.04	&	0.14	\\ 
SN2005eq	&	53654.4	&	0.1	&	0.82	&	0.01	&	35.46	&	0.07	\\ 
SN2005hc	&	53666.7	&	0.1	&	0.80	&	0.01	&	36.50	&	0.05	\\ 
SN2005hj	&	53673.8	&	0.2	&	0.80	&	0.02	&	37.03	&	0.04	\\ 
SN2005iq	&	53687.7	&	0.1	&	1.28	&	0.01	&	35.80	&	0.15	\\  
SN2005kc	&	53697.7	&	0.1	&	1.12	&	0.02	&	33.89	&	0.15	\\ 
SN2005ki	&	53705.5	&	0.1	&	1.36	&	0.01	&	34.73	&	0.10	\\ 
SN2005na	&	53740.2	&	0.1	&	1.03	&	0.01	&	35.34	&	0.08	\\ 
SN2006D		&	53757.7	&	0.1	&	1.47	&	0.01	&	33.00	&	0.15	\\ 
SN2006X		&	53786.3	&	0.1	&	1.09	&	0.03	&	30.91	&	0.08	\\ 
SN2006ax	&	53827.2	&	0.1	&	1.04	&	0.01	&	34.46	&	0.11	\\  
SN2006bh	&	53833.6	&	0.1	&	1.42	&	0.01	&	33.28	&	0.20	\\ 
SN2006br	&	53853.7	&	0.6	&	1.45	&	0.05	&	35.23	&	0.08	\\  
SN2006ej	&	53976.4	&	0.2	&	1.37	&	0.01	&	34.62	&	0.11	\\ 
SN2006eq	&	53975.9	&	0.4	&	1.88	&	0.04	&	36.66	&	0.04	\\ 
SN2006et	&	53993.7	&	0.1	&	0.88	&	0.01	&	34.82	&	0.10	\\ 
SN2006ev	&	53990.1	&	0.3	&	1.34	&	0.01	&	35.40	&	0.08	\\ 
SN2006gj	&	54000.3	&	0.2	&	1.56	&	0.04	&	35.42	&	0.08	\\ 
SN2006gt	&	54003.1	&	0.3	&	1.71	&	0.03	&	36.43	&	0.05	\\ 
SN2006hb	&	54006.0	&	0.3	&	1.69	&	0.02	&	34.11	&	0.13	\\ 
SN2006hx	&	54021.9	&	0.2	&	1.07	&	0.05	&	36.47	&	0.05	\\ 
SN2006is	&	54007.5	&	0.4	&	0.80	&	0.01	&	35.69	&	0.07	\\ 
SN2006kf	&	54041.3	&	0.1	&	1.51	&	0.01	&	34.78	&	0.10	\\ 
SN2006lu	&	54034.4	&	0.2	&	0.92	&	0.01	&	36.92	&	0.04	\\ 
SN2006ob	&	54063.4	&	0.1	&	1.51	&	0.01	&	37.08	&	0.04	\\ 
SN2006os	&	54063.9	&	0.2	&	1.08	&	0.02	&	35.70	&	0.07	\\ 
SN2007A		&	54113.1	&	0.2	&	1.06	&	0.04	&	34.26	&	0.13	\\  
SN2007S		&	54143.8	&	0.1	&	0.81	&	0.01	&	34.06	&	0.14	\\ 
SN2007af	&	54174.4	&	0.1	&	1.11	&	0.01	&	32.10	&	0.10	\\ 
SN2007ai	&	54173.5	&	0.3	&	0.84	&	0.02	&	35.73	&	0.07	\\ 
SN2007as	&	54181.3	&	0.4	&	1.27	&	0.03	&	34.45	&	0.12	\\  
SN2007bc	&	54200.3	&	0.2	&	1.27	&	0.02	&	34.89	&	0.10	\\ 
SN2007bd	&	54206.9	&	0.1	&	1.27	&	0.01	&	35.73	&	0.07	\\ 
SN2007bm	&	54224.1	&	0.2	&	1.11	&	0.02	&	32.30	&	0.07	\\ 
SN2007ca	&	54227.7	&	0.2	&	1.05	&	0.03	&	34.04	&	0.14	\\ 
SN2007if	&	54343.1	&	0.6	&	1.07	&	0.03	&	37.59	&	0.03	\\ 
SN2007jg	&	54366.1	&	0.3	&	1.09	&	0.04	&	36.03	&	0.06	\\ 
SN2007le	&	54399.3	&	0.1	&	1.03	&	0.02	&	32.34	&	0.08	\\  
SN2007nq	&	54398.8	&	0.1	&	1.49	&	0.01	&	36.44	&	0.05	\\ 
SN2007on	&	54419.8	&	0.4	&	1.65	&	0.04	&	31.45	&	0.08	\\ 
SN2008C		&	54466.1	&	0.2	&	1.08	&	0.02	&	34.34	&	0.12	\\ 
SN2008R		&	54494.5	&	0.1	&	1.77	&	0.04	&	33.73	&	0.16	\\ 
SN2008bc	&	54550.0	&	0.1	&	1.04	&	0.02	&	34.16	&	0.13	\\ 
SN2008bq	&	54562.1	&	0.2	&	0.78	&	0.02	&	35.79	&	0.06	\\ 
SN2008fp	&	54730.9	&	0.1	&	1.05	&	0.01	&	31.79	&	0.05	\\ 
SN2008gp	&	54779.1	&	0.1	&	1.01	&	0.01	&	35.79	&	0.06	\\ 
SN2008hv	&	54817.1	&	0.1	&	1.30	&	0.01	&	33.84	&	0.15	\\ 
SN2008ia	&	54813.2	&	0.1 	&	1.34	&	0.01	&	34.96	&	0.09	\\ 
SN2011fe	&	55815.0	&	0.3	&	1.20	&	0.02	&	28.91	&	0.20	\\ 
SN2014J		&	56689.7	&	0.3	&	1.10	&	0.02	&	27.64	&	0.10	\\ 
\hline
	
\end{tabular}
\label{tab:snpy}
\end{center}
\end{minipage}
\end{table}

While the extinction is much reduced in the NIR it is not entirely
negligible. Using the Cardelli extinction law \citep{Cardelli1989}, the
extinction in the $H$-band is $\sim$18\% of that in $V$-band. We have also included some
heavily-extinguished SNe\,Ia like SNe 1986G, 2005A, 2006X, 2006br
and 2014J without an extinction correction in our sample. Therefore, the observed scatter is
larger than the  intrinsic variation among SNe\,Ia. We have chosen not to apply a 
correction for the host galaxy extinction as the reddening law remains under 
debate \citep[e.g.][]{Phillips2013}.

We fit a spline interpolation to the data to derive the phase and magnitude at maximum, the minimum
and the second maximum in each filter. In order for a measurement of the minimum and second maximum to be made, we require $\geq$4~observations at late phases 
($>$7 d for the minimum and $>$15 d
for the second maximum). We also require that observations at
least 4~d before $t_{max}$ were available. The uncertainties for each derived parameter were
calculated by repeating the fits to 1000 Monte Carlo realisations of light curves 
generated using the errors on the photometry.

The NIR light curves are very uniform up to the time of the $B$-band maximum, $\sim$3--4 d after the maximum is reached in the NIR light
curves. In the lower panels in Fig.~\ref{fig:lc1}, we show the RMS scatter
for each epoch.

The scatter remains small for $\sim$1 week around maximum in the different bands independent
of the sample (CSP or non-CSP). The two samples show very
consistent scatter out to a phase of $\sim$10~d in $J$ and $\sim$25~d in $H$, 
after which they start to deviate. With only 14 objects
in the non-CSP sample in this phase range compared to 25 from the CSP, we 
consider the differences not statistically significant.  
The scatter 
continues to increase beyond 35 d, which we attribute to the colour evolution (see \S\ref{sec:cols}
below).

\subsection{The first maximum}
\label{sec-first}

\begin{figure}
\includegraphics[width=0.45\textwidth, height=0.5\textheight]{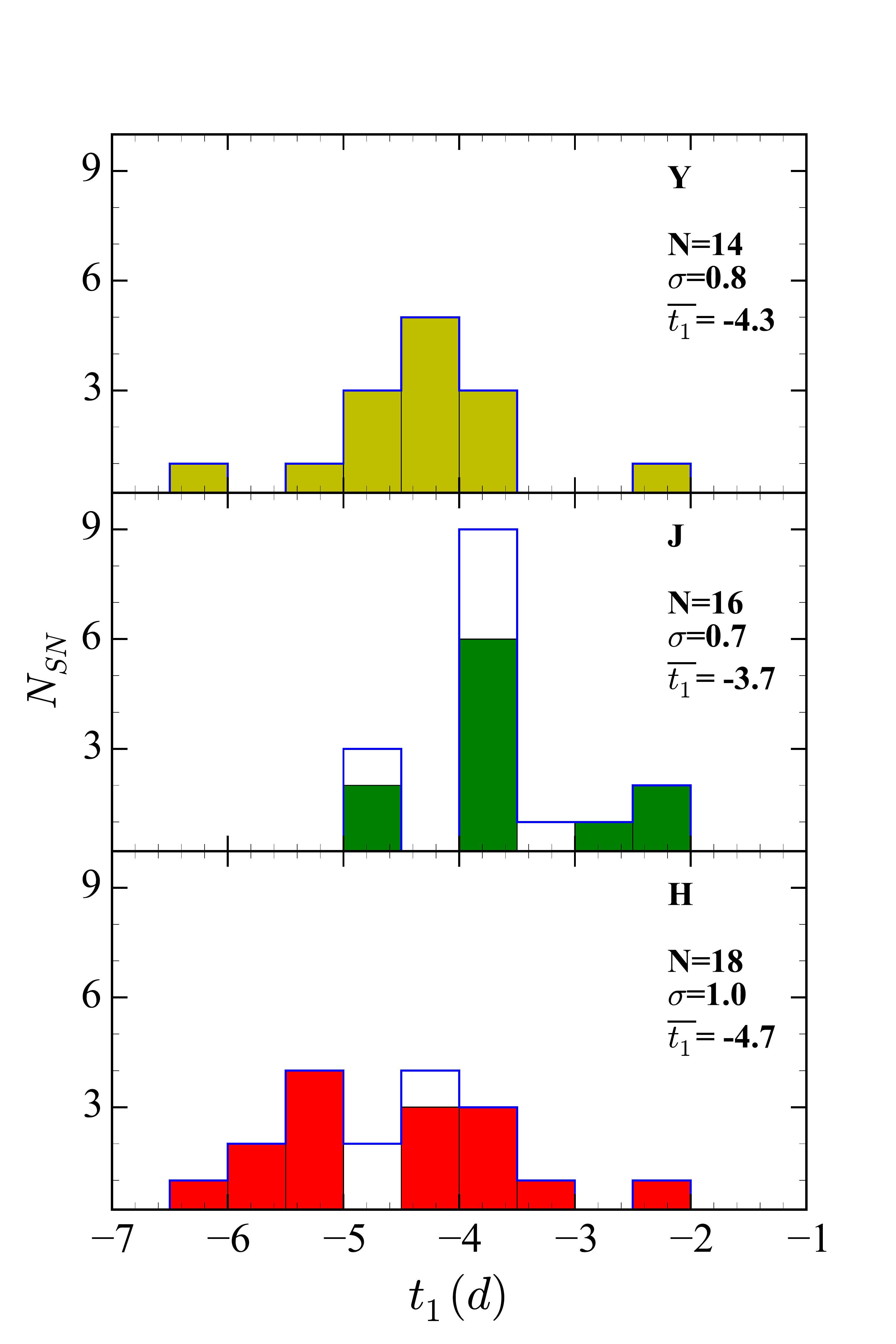}
\caption{
Distribution of $t_1$ for the $Y$, $J$, and $H$ light
curves relative to the $B$-band maximum. The filled histograms are for
SNe\,Ia from the CSP sample while the open histograms show the combined CSP and non-CSP sample. In the $Y$ band, we only use the objects observed by the CSP.
 The NIR
light curves peak at within $-2$ to $-7$ d relative to the
$B$ maximum.
}
\label{fig:hist_t1}
\end{figure}

\cite{Elias1981} showed, in a small sample of SNe Ia, that the
$JHK$ light curves of SNe\,Ia peak earlier than the optical light curves.
We confirm this for our sample (Fig.~\ref{fig:hist_t1}).   

The NIR light curves peak
within $-2$ to $-7$ d of the $B$-band peak confirming the result of
\citet{Folatelli2010} for SNe\,Ia with $\Delta m_{15}(B)$$<$1.8. There is
no obvious difference between our full sample and the CSP sample as seen
in the scatter. The distribution of $t_1$ is remarkably tight
(less than one day dispersion) for all filters. This indicates a close
relationship between $t^{max}_{B}$ and the NIR $t_1$ values for the SNe\,Ia in our sample. 

Table~\ref{tab:scat} gives the phase of lowest scatter measured in
each filter. Without any attempt to normalise the light curves, we find
the smallest scatter in all NIR light curves near $t_1$.
The dispersion remains very low for $\sim$1 week, before increasing to
$>$0.2 mag at later phases.

\subsection{The minimum}
\label{sec-min}

\begin{figure}
\includegraphics[width=0.45\textwidth, height=0.5\textheight]{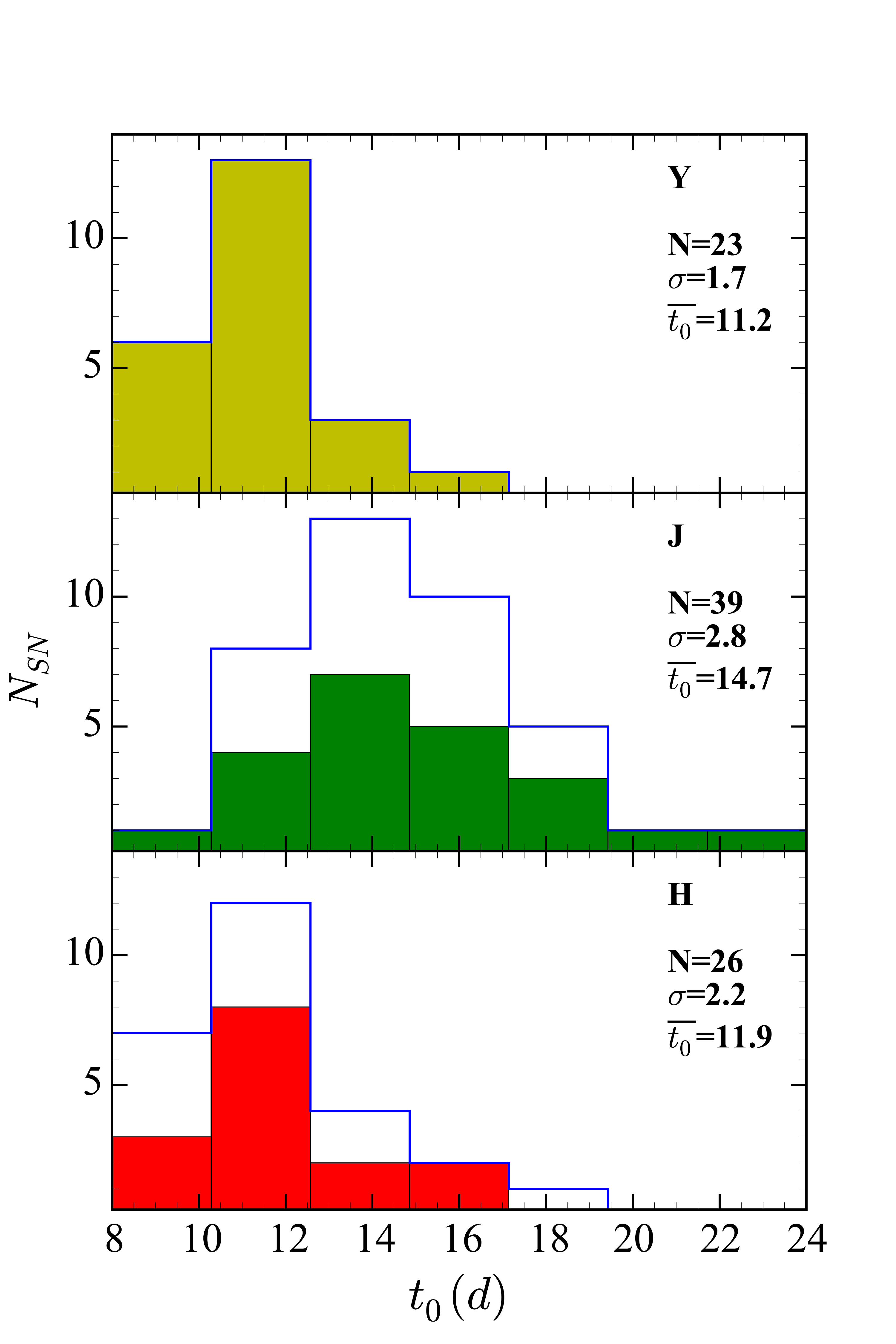}
\caption{
Distribution of $t_0$ in $Y$, $J$, and $H$ light curves.
The $Y$ and $H$ minimum is reached a few days before the $J$ light curve
dips, which occurs $\sim$2 weeks after $B$-band maximum.  
}
\label{fig:hist_t0}
\end{figure}

The minimum in $J$ occurs $\sim$2 weeks after $t^{max}_{B}$
(Fig.~\ref{fig:hist_t0}). The $Y$ light curves dip about 3 d
earlier at $t_0=+11$\,d. The minimum in $H$ is reached on average
about 2 d before $J$ at $t_0=+12$\,d. The phase range is still
relatively narrow with the minima all occurring within roughly $\pm
2$\,d. While $Y$ and $H$ display a tight distribution of t$_0$, 
the $J$ distribution exhibits a tail of late minima. 

\begin{table}
\small
\centering
\caption{Magnitude scatter in NIR light curves. The second column
indicates the time of minimum magnitude scatter in each filter, while the third
column gives the scatter at this phase. The fourth column gives the
phase range for which the scatter stays below 0.2 mag.}
\begin{tabular}{l c c c l}
\hline
Filter  & $t$  & $\sigma(M)$ &	Phase range & SN sample \\
        & (days)      &    (mag)    & ($\sigma(M) < 0.2$~mag)  & \\
\hline
Y	& $-4.4$      &       0.15  &   [$-4$ , +1] &	CSP\\
\hline
J	& $-3.6$      &       0.16  &   [$-4$ , +3] &	CSP\\
J	& $-3.8$      &       0.17  &   [$-6$ , +1] &	non-CSP \\
\hline
H	& $-5.1$      &       0.17  &   [$-5$ , +1] &	CSP\\
H	& $-4.7$      &       0.14  &   [$-7$ , +2] &   non-CSP\\
\hline
\end{tabular}
\label{tab:scat}
\end{table}

No significant difference between the CSP and the literature samples can
be seen in the distributions. The scatter among $M_0$ is fairly small for the three NIR bands although about 2 to 3 times larger than at t$_1$.
In the $Y$-band, the scatter is low ($<$0.2 mag) immediately after t$_0$, while the $J$ and $H$ light
curves display a larger dispersion at this point.

\subsection{The second maximum} 
\label{ssec-smax}

\begin{figure}
\includegraphics[width=.45\textwidth, height=0.5\textheight]{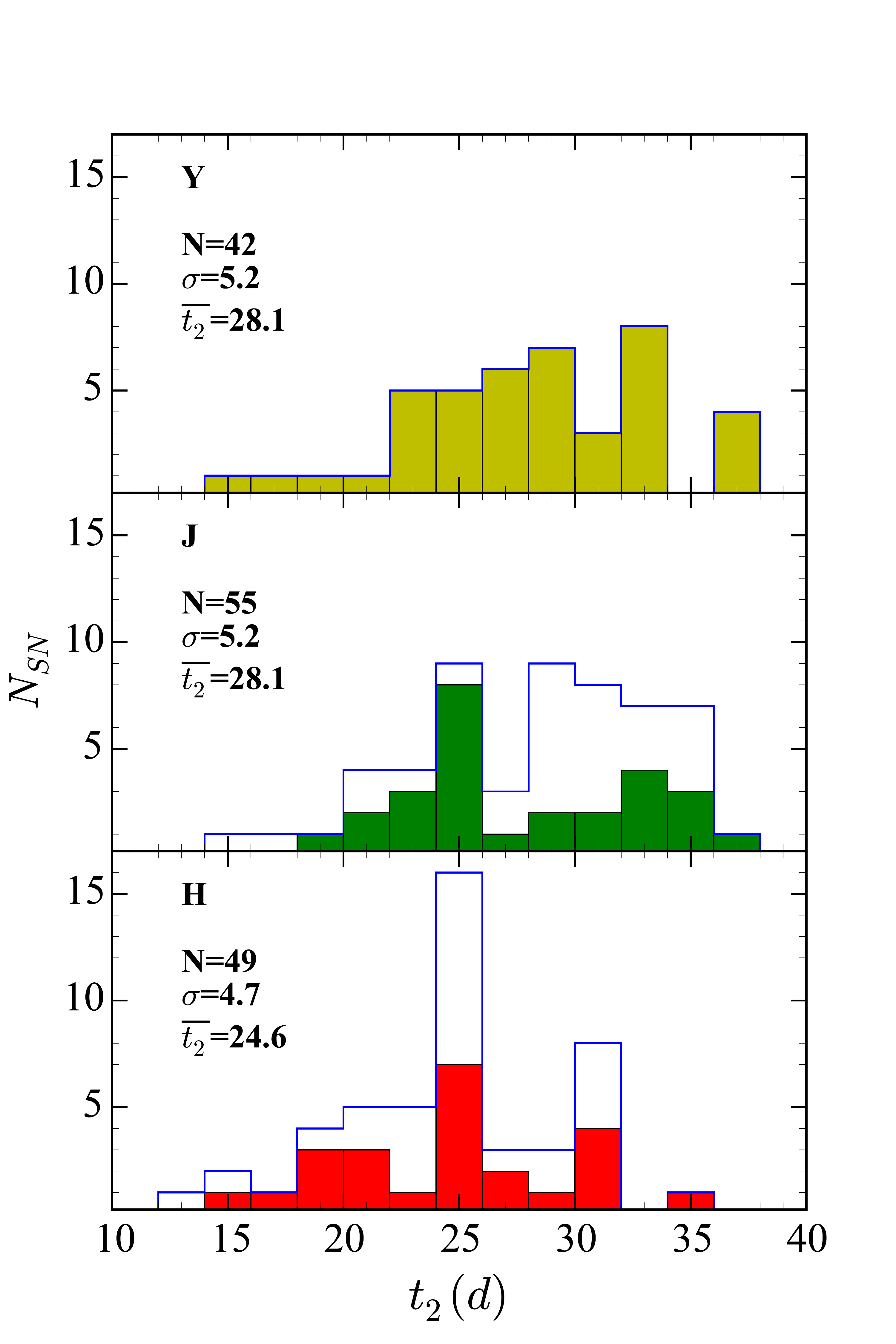}
\caption{
Distribution of t$_2$ in the $Y$, $J$ and $H$ light curves. Note
the expanded scale on the abscissa compared to the phase ranges in
Figs.~\ref{fig:hist_t1} and \ref{fig:hist_t0}. 
}
\label{fig:hist_t2}
\end{figure}

Figure~\ref{fig:hist_t2} shows that t$_2$ occurs over a
wide range of phases and can vary by as much as 20 d from one SN\,Ia
to another. This diversity had been observed before for the $i$ 
light curves
\citep[e.g.][]{Hamuy1996, Folatelli2010} and for $JHK$
\citep{Mandel2009, Biscardi2012}. In Figure~\ref{fig:hist_t2}, we show
that the mean t$_2$ is later in $Y$- and $J$-bands, with the $H$-band light curves reaching t$_2$, on average, a few
days earlier. 
The scatter in the luminosity starts to increase slowly after t$_0$, and is seen to increase significantly ($>$0.5 mag) around the time of t$_2$  (Fig.~\ref{fig:lc1}).

\subsection{The late decline}
\label{sec-late}

\begin{figure}
\centering
\includegraphics[width=0.45\textwidth, height=0.5\textheight]{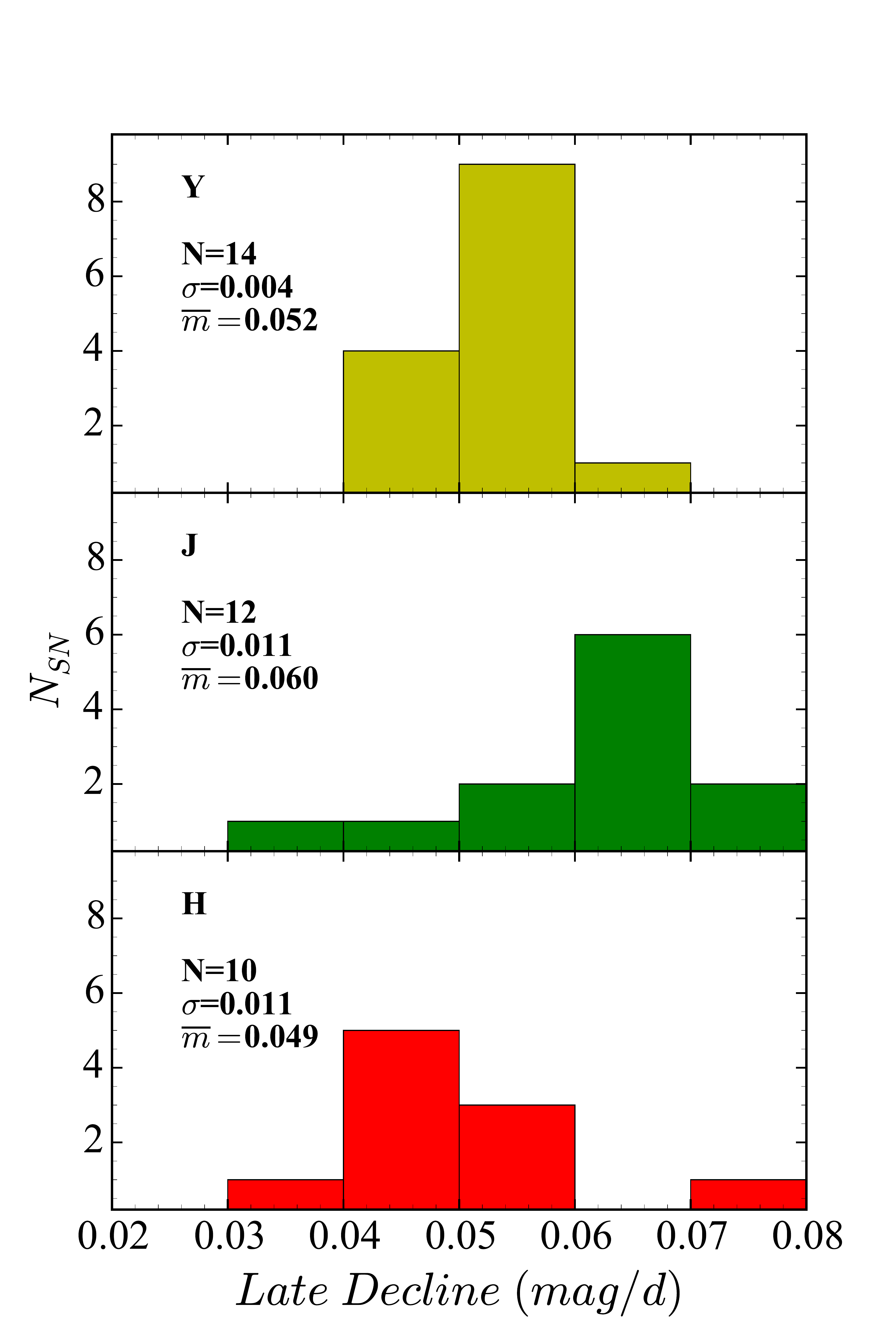}
\caption{
Distribution of decline rates after t$_2$.
}
\label{fig:late}
\end{figure}

After the second maximum, the light curves steadily decline. In Fig.~\ref{fig:late}
we show the distribution of slopes calculated for SNe Ia with at least
three observations in phases 40$<$$t$$<$90 d. This choice of phase range ensures that the measurements are not influenced by the second maximum. All the SNe Ia in our sample with data at these late phases come from the CSP.

There is a tight distribution of decline rates in the $Y$- and $H$-bands, with the notable
exception of SN~2005M in $H$, which declined twice as fast as the other
SNe Ia.  The scatter in the $H$-band decline rate, when excluding SN~2005M, is
only 0.004 mag per day, similar to the scatter in the $Y$ filter.
The same SN is also one of the two fast declining objects in $J$
(in the fastest bin, \emph{middle panel}), where the scatter in the decline rates is significantly larger than for the other two bands. 

\subsection{NIR colours}
\label{sec:cols}

\begin{figure*}
\includegraphics[width=0.8\textwidth, height=0.4\textheight]{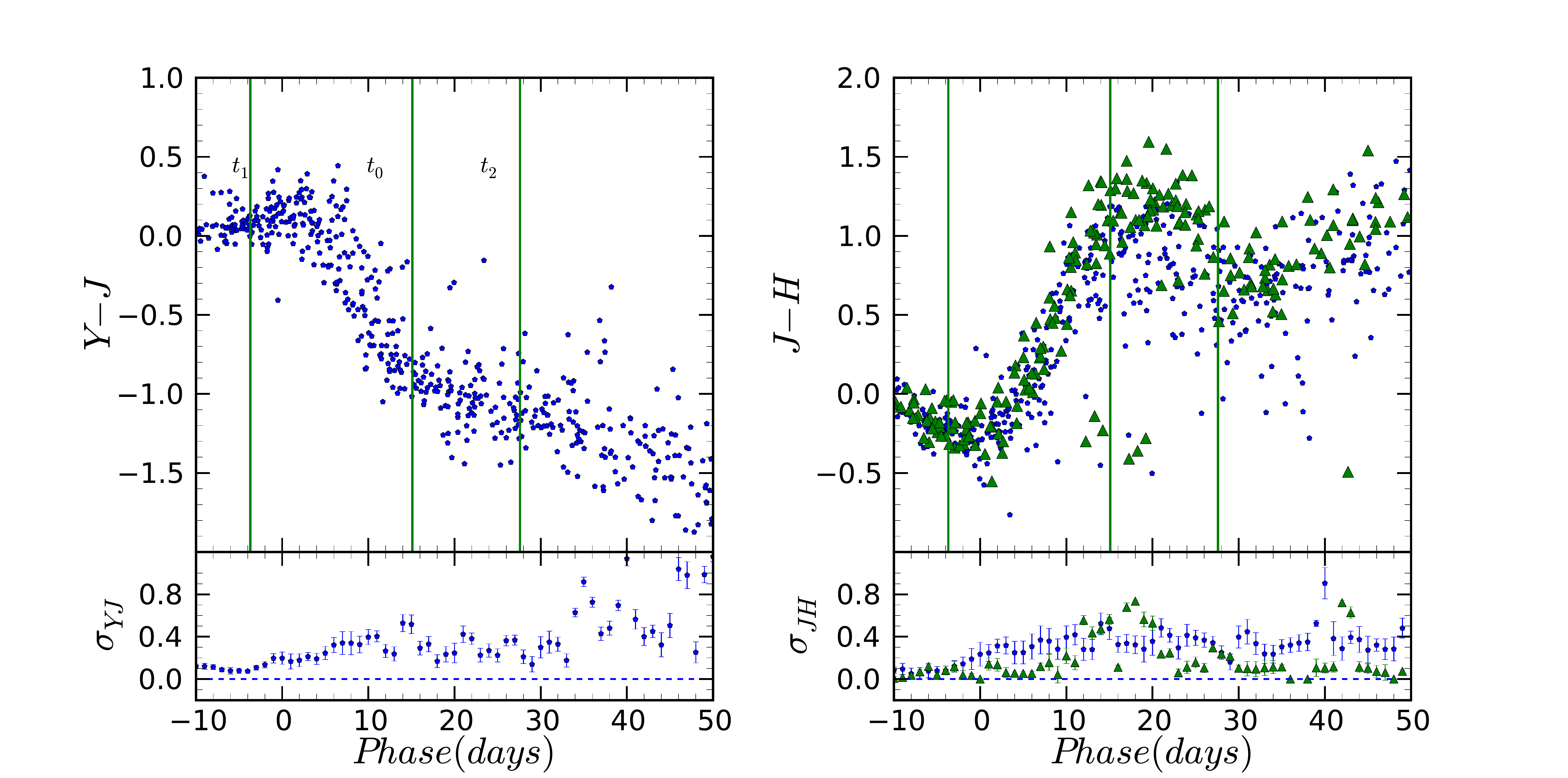}
\caption{
\emph{Left}: $Y-J$ colour curve. The mean $t_1$,
$t_0$ and $t_2$ in the $Y$ filter are plotted as green lines. The scatter plot
in the lower panel only contains information from the CSP objects since
the non-CSP sample does not have $Y$-filter observations. The blue points are the objects in the CSP sample, whereas the
green points are the non-CSP objects. \emph{Right}:
$J-H$ colour curve. The epoch of minimum scatter in $J$, the average
value of $t_0$ and $t_2$ in $J$ are overplotted in green. The lower
panel shows the evolution of the scatter around the mean in the colour
curve. 
}
\label{fig:col_scat}
\end{figure*}

\cite{Elias1985} showed that the early $J-H$ colour
evolution is rather uniform for SNe\,Ia. In
Fig.~\ref{fig:col_scat}, we show the colour curves ($Y-J$ and $J-H$) for the CSP and
non-CSP objects. The scatter at each epoch is plotted in the lower
panels. Similar to the light curves, the early colour evolution is similar for
most SNe\,Ia in our sample. 

At the first maximum the scatter is minimal in $Y-J$
($\sigma(Y-J) = 0.07$~mag) and in $J-H$ ($\sigma(J-H)=0.05$~mag). The scatter stays $<$0.1 mag between $-7$ and $-3$
d for $Y-J$ and between $-10$ and $-3$ d for $J-H$. At early times, we find
that the 
samples display the same scatter.  Immediately
after the first maximum, the colour curves 
start to deviate like the
light curves. 
By the time of the minimum the colours display a wide
spread which continues to increase into the late decline.

The $Y-J$ colour remains fairly constant until about one week after
maximum when the $Y-J$ colour starts a monotonic evolution towards bluer colours. For the
$J-H$ colour, SNe\,Ia evolve to redder colours after maximum. Then after the light
curve minimum, the $J-H$ colour tends towards slightly bluer colours until
$\sim$30 d.

\section{Correlations}
\label{sec-corr}
 
The uniformity of the NIR light curves at
maximum light suggests that the size of the surface of last scattering
at these wavelengths is independent of the details of the explosion
and  progenitor. 

A diversity in the NIR only becomes visible at later phases as the lines contributing to the
line blanketing opacity arise from deeper in the ejecta. The 
phase and magnitude the minimum
and in particular the second maximum of the NIR light curves display
large variations. Correlating these changes with 
other SN
parameters should shed light on the physical processes underlying the
explosions and the release of the radiation. 

\begin{figure}
\centering
\includegraphics[width=0.45\textwidth, height=0.5\textheight,
trim= 0 30 0 30]{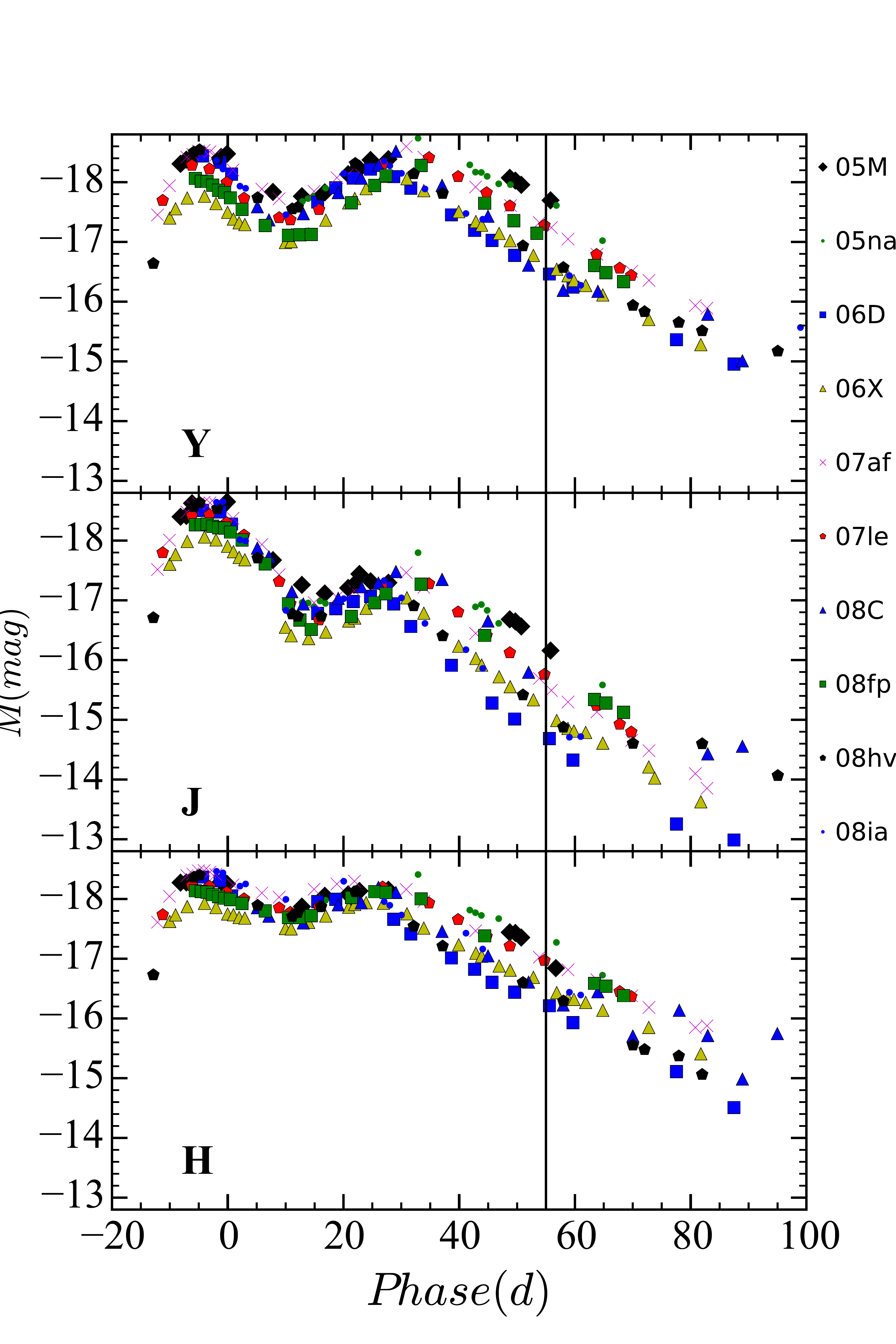}
\caption{
Complete absolute magnitude light curves in $YJH$ for the 10 objects for which a late
decline is measured in all three filters. The figure illustrates that
the decline rates at late times are uniform, whereas there is a large
scatter in the absolute magnitude at +55 d indicated by the vertical
line.
}
\label{fig:lc_late}
\end{figure}

In Fig.~\ref{fig:lc_late} we show example NIR light curves
extending to late times for 10 SNe Ia in our sample. It should be noted that
among these 10 SNe are two where the extinction has been determined to
be high (SN~2006X and SN~2008fp). Therefore, some of the scatter can be
attributed to uncorrected host galaxy extinction.

In the following sections, we note that correlations reported with $r > 0.4$ are significant, and those with $r> 0.65$ are termed as strong.   

\subsection{NIR light curve properties}
\label{sec:ir-lc}

In the previous section, we found a large diversity in the NIR light
curve properties at post-maximum epochs. In this section, we investigate
correlations between these properties in more detail. We find a correlation between
$t_0$ and $t_2$ in $JH$ filters (Pearson parameter $r \sim 0.5$) suggesting that a later minimum is followed by a
later second maximum. We also find a correlation between $M_2$ and $M_0$ in $Y$ and $J$  filters, with $r=0.59$ and $r=0.50$ respectively. This implies that SNe\,Ia with a more luminous minimum also display a more
luminous second maximum.

Interestingly, the $M_2$ and $t_2$ do not correlate ($r < 0.4$) in any filter. However, there we do find a significant correlation across different filters between $M_2$ and $t_2$ (e.g. $M_2$  in $Y$ band correlates with $t_2$ in $J$ band with $r=0.61$).
\citet{Mandel2009} found that the 
rate of luminosity increase (rise rate $r/\beta$ in the \citealt{Mandel2009} nomenclature)
to the second maximum
correlated with the luminosity of the first peak. We cannot confirm this
at any significance with our data. 

Comparison of Figs.~\ref{fig:hist_t0} and \ref{fig:hist_t2} reveals that
the light curve evolution in the $Y$-band is slowest amongst the NIR filters.
SNe~Ia on average reach t$_0$ in $Y$ earlier than in $J$ and $H$,
but reach t$_2$ later in the $Y$ compared to the other filters. The rise time in $Y$
is nearly 4 d longer than in $J$ and nearly 7 d longer
than in $H$. 

\begin{figure}
\centering
\includegraphics[width=.5\textwidth, height=0.33\textheight]{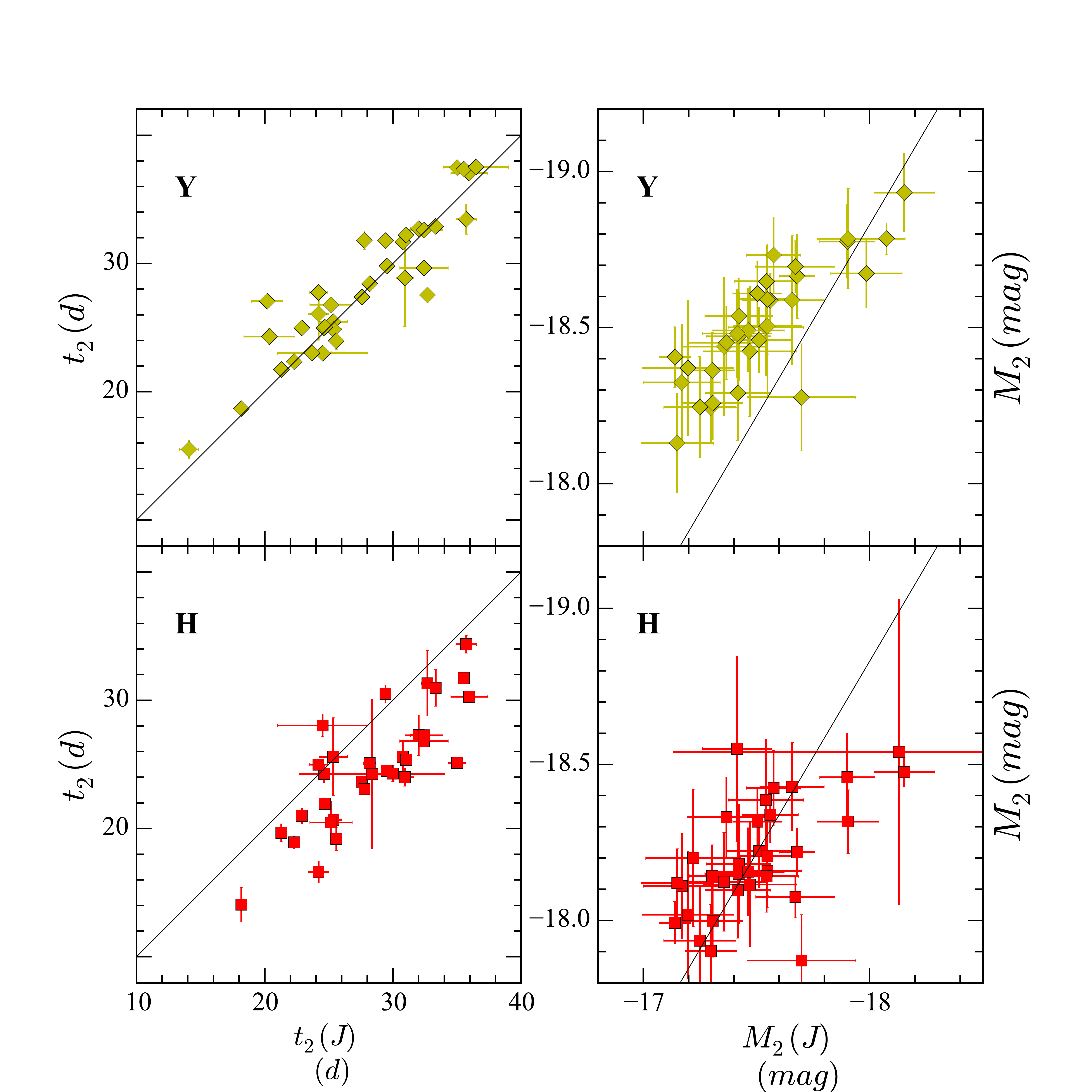}
\caption{
Phases (\emph{left}) and luminosity (\emph{right}) of the second maximum
in NIR filters. There are clear correlations between the
filters with the weakest trend in the $H$ versus $J$ luminosities. The black line is a one-to-one relation.
}
\label{fig:filt_m2}
\end{figure}

The phase and luminosity of the second maximum  strongly correlate
between the NIR filters (Fig.~\ref{fig:filt_m2}). SNe~Ia with a later t$_2$ display a higher luminosity during the late decline. The luminosity at 55 d
past the $B$ maximum, hereafter referred to as $M|_{55}$ was
chosen to ensure all SNe~Ia  have entered the late decline well past t$_2$. 
A choice of a later phase may be more representative of the decline but would
result in a smaller SN sample as not many objects are observed
at these epochs and the decreasing flux results in larger uncertainties.

\begin{figure}
\centering
\includegraphics[width=.40\textwidth, height=0.5\textheight]{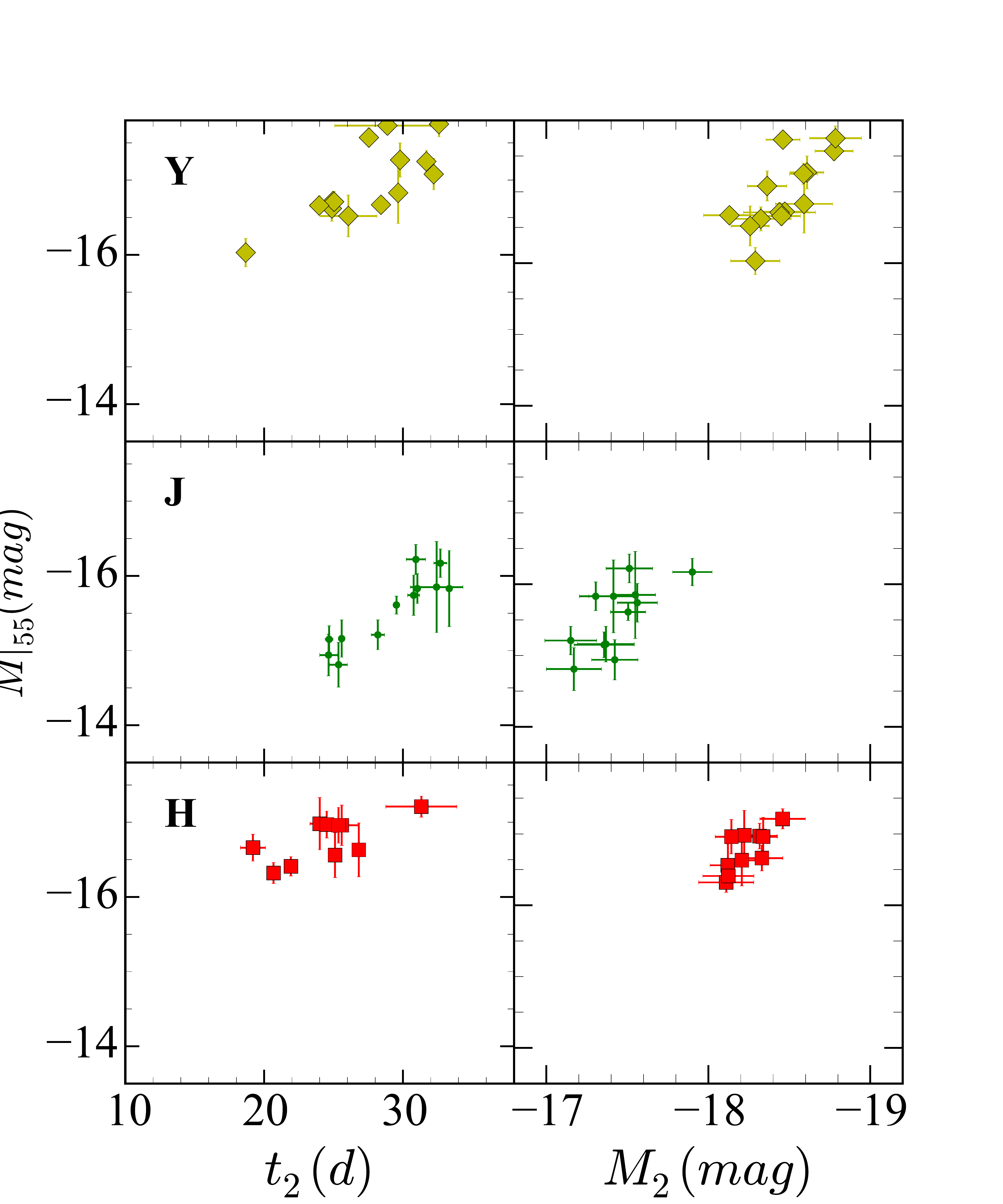}
\caption{
(\emph{Left:}) Absolute magnitude at $t=55$~d in $YJH$ {\em vs.}
$t_2$. (\emph{Right:})
$M|_{55}$ compared to the absolute magnitude of the second maximum $M_2$.
}
\label{fig:j55}
\end{figure}

In Figure \ref{fig:j55} we plot $M|_{55}$ against $t_2$ and $M_2$. A
clear trend between $M|_{55}$ and $t_2$ is present in all filters
(Pearson coefficients $r$
of 0.78, 0.92, 0.68 in the $YJH$, respectively;
Fig.~\ref{fig:j55}, left panels). At this phase SNe~Ia are on average
most luminous in $Y$ followed by $H$ and $J$, a trend that is already
present at $M_2$. This is also reflected in the NIR colour evolution (see section~\S\ref{sec:cols}).

The dispersion in $M|_{55}$ is large with $\sigma(M|_{55})=0.48$~mag,
0.51~mag and 0.30~mag in $Y$, $J$ and $H$, respectively.
This is not unexpected as it continues the trend to larger 
(luminosity) differences in the NIR light curves with increasing
phase.

\subsection{Correlations with optical light curve shape parameters}
\label{sec:opt}

It is interesting to see whether the NIR light curve parameters
correlate with some of the well known optical light curve shape features 
\citep[$\Delta m_{15}$;][]{Burns2011}. 

\citet{Folatelli2010} have shown that the value of
$t_2$ in $i$ correlated with $\Delta m_{15} (B)$.
Since the dispersion
increases with phase we concentrate on the second maximum and explore
whether the timing and the strength correlate with the optical light curve shape. 

The second maximum in the NIR is a result of an ionization transition of
the Fe-group elements from doubly to singly ionized states.
Models
predict that $t_2$ depends on the amount of
Ni ($M_{Ni}$) synthesized in the explosion \citep{Kasen2006}.  Since
$M_{Ni}$ is known to correlate with $m^{max}_{B}$, which itself is a function of the light curve shape
\citep{Arnett1982, Stritzinger2006, Mazzali2007, Scalzo2014}, we
explored the relation between $t_2$ and $\Delta m_{15}$
(Tab.~\ref{tab:snpy}).  

\begin{figure}
\centering 
\includegraphics[width=.47\textwidth, height=0.5\textheight]{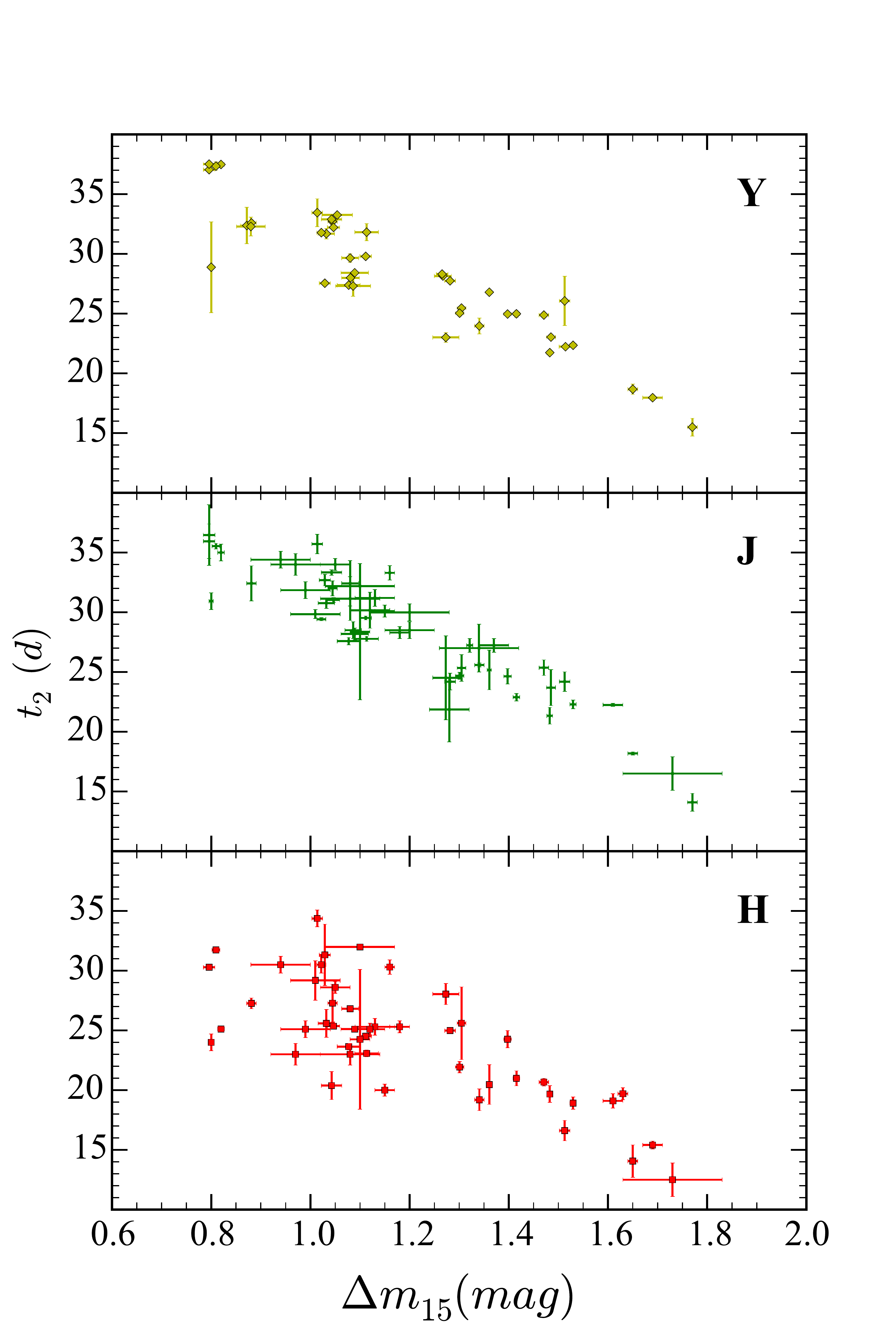} 
\caption{
Comparison of t$_2$  in the NIR light
curves with $\Delta
m_{15}(B)$. 
}
\label{fig:t2} 
\end{figure} 

Figure \ref{fig:t2} confirms a strong correlation between $t_2$ and
$\Delta m_{15}$. The Pearson coefficients are 0.91, 0.93
and 0.75 for the $Y$, $J$, and $H$-bands, respectively. The weaker
correlation in $H$ may be due to a low dependency on t$_2$ for objects with $\Delta m_{15}$ $<$ 1.2, where $t_2$
appears independent of $\Delta m_{15}$. This applies only to SNe\,Ia
with a slow optical decline. For higher $\Delta m_{15}$, the trend is as
strong as in the other filters. 

A linear regression yields equation~\ref{eq:lin_t2} for each filter. The
RMS scatter is 2.1 d, 1.8 d and 3.0 d in the $YJH$ filters. 

\begin{subequations}
\label{eq:lin_t2}
\begin{equation}
\label{eq:y}
t_2(Y)=(-20.6 \pm 1.0) \Delta m_{15}+(53.6 \pm 1.2)
\end{equation}
\begin{equation}
\label{eq:j}
t_2(J)=(-20.3 \pm 0.9) \Delta m_{15}+(51.7 \pm 1.1)
\end{equation}
\begin{equation}
\label{eq:h}
t_2(H)=(-16.0 \pm 2.7) \Delta m_{15}+(42.3 \pm 2.8)
\end{equation}
\end{subequations}

Since the phase of the second maximum ($t_2$) in NIR bands 
is strongly correlated with $\Delta m_{15}$ and thereby the optical maximum luminosity, $t_2$ could also serve as an indicator of the luminosity of
SNe\,Ia. It is noteworthy that even extreme cases, like SN\,2007if, which
have been associated with super-Chandrasekhar mass progenitors
\citep{Scalzo2010, Scalzo2012} are fully consistent with the derived
relations. 

\begin{figure}
\centering
\includegraphics[width=.47\textwidth, height=.50\textheight]{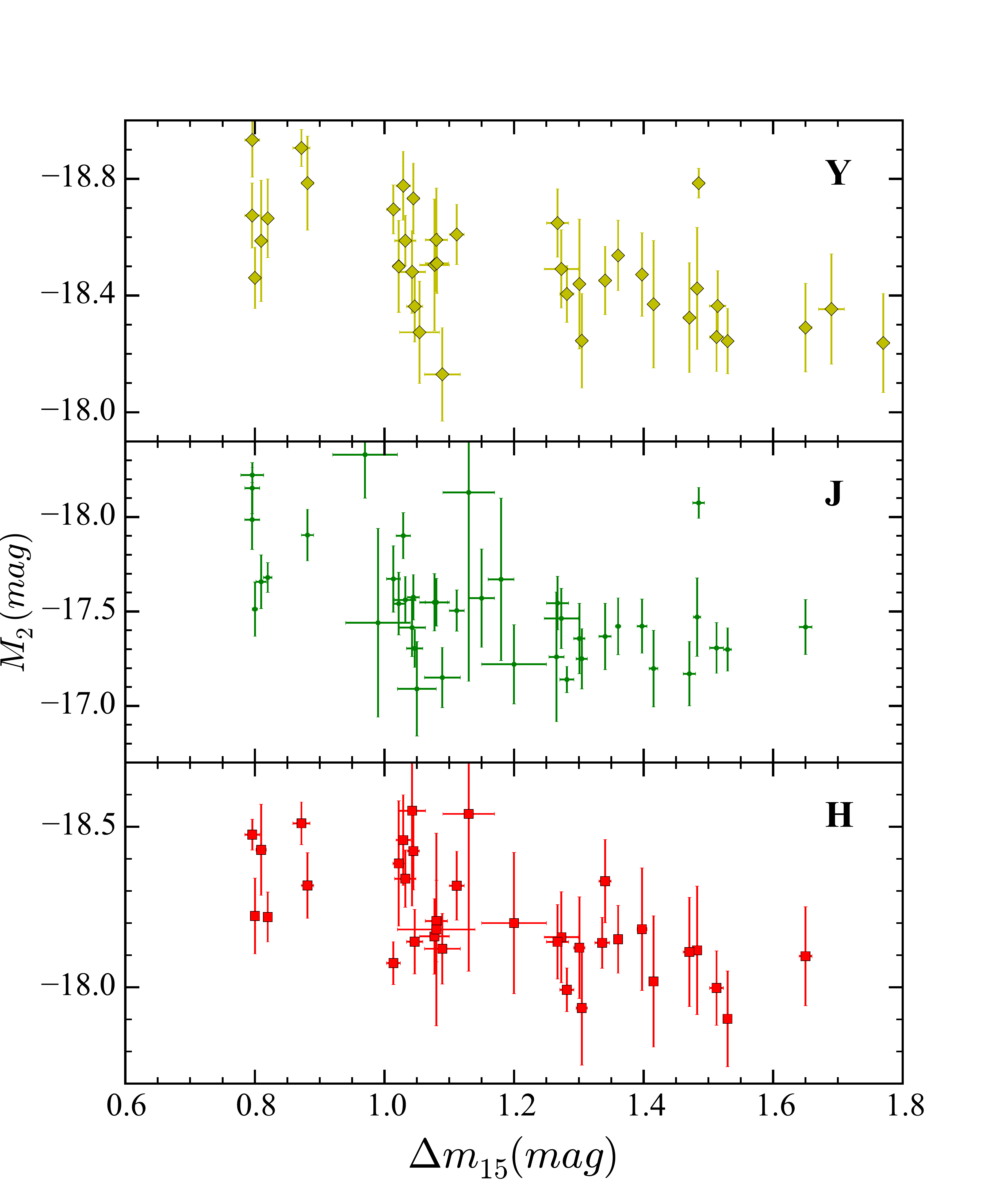}
\caption{
$M_2$ versus $\Delta m_{15}$ is shown and a weak
correlation is observed. \label{fig:m2}  
}
\end{figure}

There is a weak correlation between $M_2$ and $\Delta m_{15}$ as shown
in Fig.~\ref{fig:m2}, with very little difference between objects. 
We find $r$ values of 0.59, 0.50 and 0.63 for the
$Y$, $J$, and $H$ filters, respectively.

\section{Discussion}
\label{sec-disc}
As has been shown earlier in this paper and in a number of other
publications \citep{Meikle2000,WV08,BN12,Weyant2014}, SN\,Ia NIR light
curves show remarkable uniformity around maximum light, compared to at optical wavelengths.
This uniformity, combined with reduced
effects of extinction, holds great promise for the use of SNe\,Ia as distance indicators
in the NIR. \citet{Kattner2012} proposed to further reduce the scatter in the luminosity of
the first maximum in the NIR by a decline rate correction similar to
the procedure in the optical. 

The spread in optical luminosity is attributed to different masses, or distributions,  of
$^{56}$Ni within the ejecta \citep{Arnett1982, Stritzinger2006,
Scalzo2014}. The indifference of the NIR maximum light to the nickel mass suggests
that it is intermediate-mass elements that dominate the opacity in these
bands at maximum light. 

The NIR light curves showed an increased dispersion at later times.
We attribute this increased scatter to differences in the speed of the
evolution of the SNe\,Ia. The phase of the second maximum depends on the mass and
distribution of $^{56}$Ni, the change in opacity, the ionisation and
the dominant species setting the emission.

Not all SNe\,Ia display a second maximum \citep[e.g.][]{Krisciunas2009}
and we restricted our analysis only to objects in which the second
maximum is defined well enough to be fit. This translates into a
sample including only SNe\,Ia with $\Delta m_{15}(B) < 1.8$ mag. 
Events without a second maximum tend to be of low luminosity, often
similar to SN\,1991bg, and with large $\Delta m_{15}(B)$. 

The strength of the second peak in the $iJHK$ light curves does not
correlate with $\Delta
m_{15}$, but the phase does \citep{Folatelli2010,Biscardi2012}. 

As presented in the previous section, we confirm this relation between
$t_2$ and $\Delta m_{15}$ for the $YJH$ filters. The correlation of
$M_2$ with $\Delta m_{15}$ is rather weak (Fig.~\ref{fig:m2}) although
it appears somewhat stronger than in \citet{Biscardi2012} (the Pearson coefficients in our data are 0.5 and 0.63 for $J$ and $H$ compared to 0.12 and 0.08 in \citealt{Biscardi2012}).

\subsection{A possible physical picture}
\label{sec:poss}

The various features of the NIR light curves can be assembled into a
physical picture of the explosions. The striking similarities of the
late decline rates, when the SN becomes increasingly
transparent to the $\gamma-$rays generated by the radioactive decays,
indicate that the internal structure of the explosions is probably
similar for all SNe\,Ia considered here. The uniform decline rates are
consistent with the predictions of \citet{Woosley2007} for a range of
Chandrasekhar-mass models, with different $M_{Ni}$ but similar
radial distribution of iron-group elements. We find that the late-time
decline rate in $J$ is faster than in $Y$ and $H$, a trend also seen
between the simulated $J$ and $H$ light curves in \citet{Dessart2014}.
The NIR light curves depend very little on the explosion geometry
\citep{Kromer2009}. The DDC15 models of \citet{Blondin2013} show that
the $YHK$ decline
rates are similar to the pseudo-bolometric decline rate (S.~Blondin,
private communication). The $J$ band shows a faster decline due to a
lack of emission features \citep{Spyr1994}.  This also explains
the evolution of the $J-H$ colour curve to redder colours at late times.

At the these late times $M|_{55}$ shows a large scatter (cf.
Fig.~\ref{fig:j55}). If the similar late decline rates ( cf. Fig. ~\ref{fig:late})  indicate a similarity in the evolution of the $\gamma$-ray escape fraction for different SNe~Ia,then, a higher luminosity would translate into a higher energy input at late phases, i.e.
a larger mass of Fe-group elements. 

\begin{figure}
\label{fig:tlr}
\centering
\includegraphics[width=.33\textwidth, height=.67\textheight]{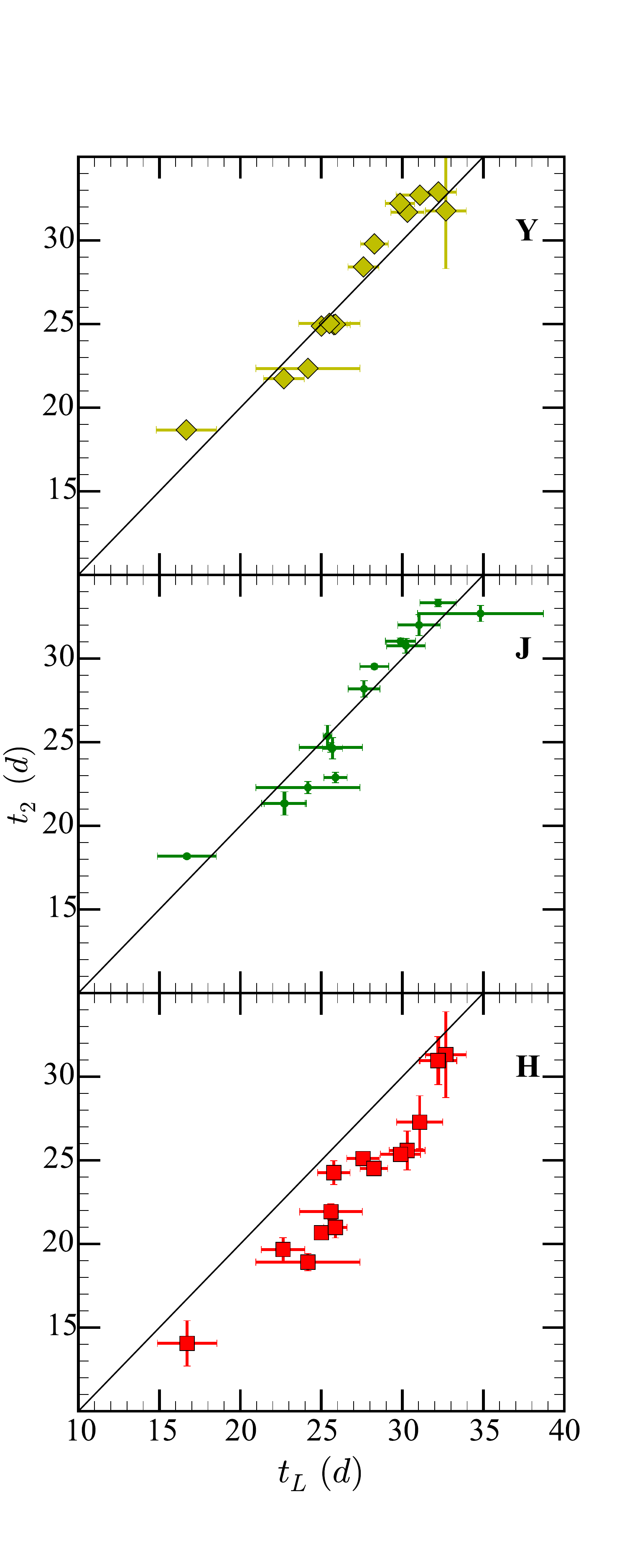}
\caption{
The phase of the second maximum versus $t_L$.
The black line is the one-to-one relation.}
\label{fig:lira}
\end{figure}

 \begin{figure*}
\includegraphics[width=1.00\textwidth, height=.43\textheight]{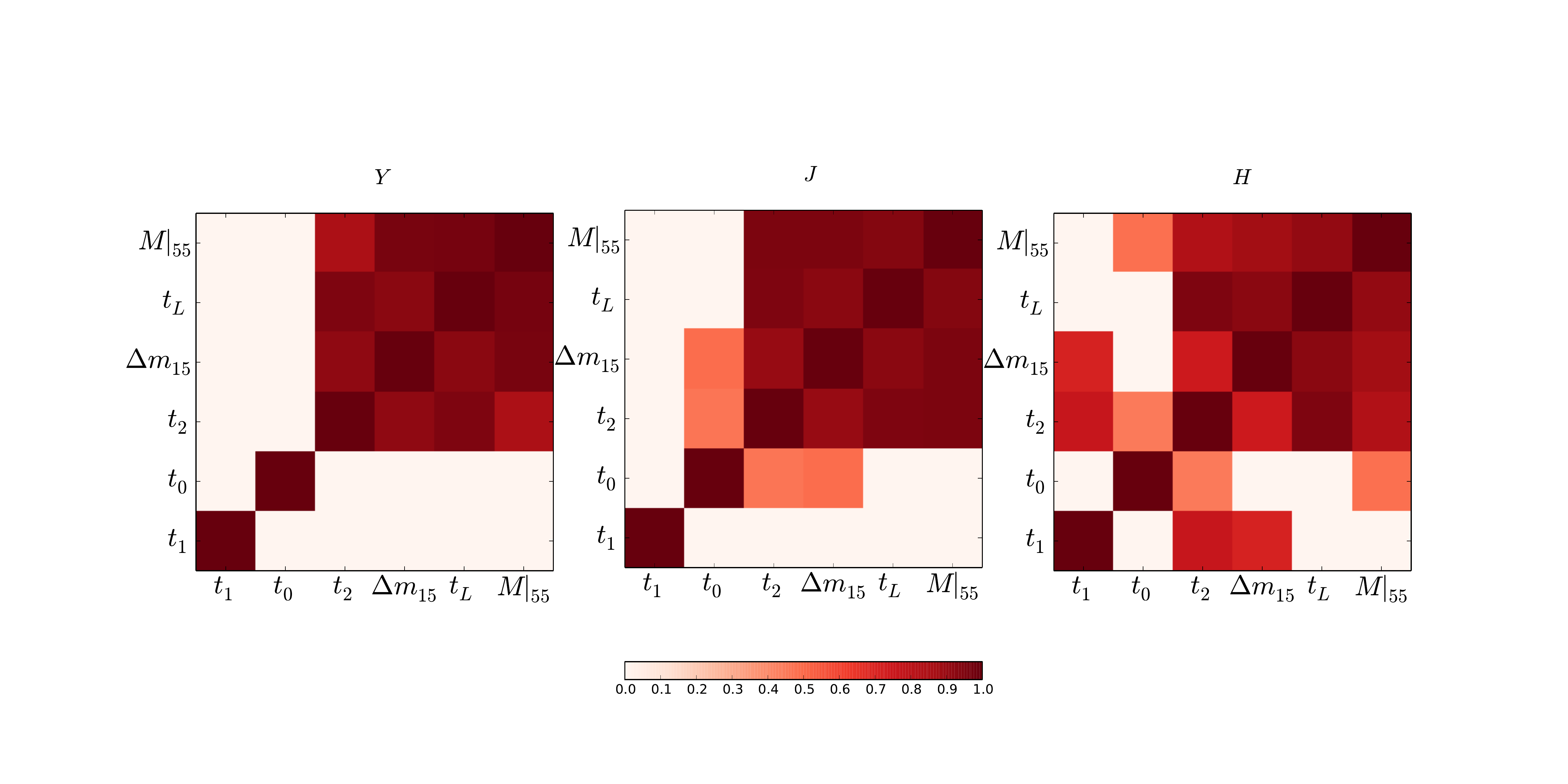}
\caption{A summary of the correlations between the timing parameters 
($t_1$, $t_0$ and $t_2$), $\Delta m_{15}$ , $t_L$ and the late time luminosity ($M|_{55}$), in each filter There is a strong correlation amongst the $t_2$, $\Delta m_{15}$, $t_L$ and $M|_{55}$ ($r> 0.65$). The colorbar scales from white to red in ascending order of correlation strength. Correlations with $r<0.4$ have been set to white
}
\label{fig:sum}
\end{figure*}
 
\citet{Kasen2006} predicted that the second maximum should be delayed
for larger Fe masses, which is exactly what is observed in the NIR
\citep[see also][]{Jack2012}. According to \citet{Kasen2007}, the faster
decline in the $B$-band light curve is mostly due to line blanketing through
Fe and Co lines, which shifts the emission into the NIR and shapes the
NIR light curves after maximum. The optical colour evolution  post $B$-maximum is suggested to be more
rapid for explosions with lower Ni masses. If this is true then the
onset of the uniform $B-V$ colour evolution \citep[referred to as the 'Lira law' and 
originally defined as the uniformity of the slope of the colour curve from 30 to 90 days, although the onset is generally at earlier phase;][]{Phillips1999} marks the beginning of the nebular phase. At these
epochs, the SN Ia spectrum is dominated by emission lines from Fe-group
elements and the emission line strength depends on the Fe mass in the
explosion. We measure the time at which the SN enters the Lira law,
hereafter $t_L$, as the epoch of inflection in the $(B-V)$ colour curve,
at late times. The procedure for measuring $t_L$ is identical to the
measurement of $t_2$ described in \S\ref{ssec-smax}.

Figure~\ref{fig:lira} shows that $t_2$ coincides nearly exactly with $t_L$ for the $Y$ and $J$ bands. While the $H$ light curve peaks 3--4
d earlier. The reason for this is not entirely clear, but both $Y$ and $J$-bands are expected to be strongly influenced by Fe lines, while $H$ is
dominated by Co lines \citep[e.g.][]{Marion2009, Jack2012}. The striking
coincidence of $t_2$ and $t_L$  in the NIR light curves is further
evidence that $t_2$ directly depends on the
Ni mass in the explosion.

\citet{Scalzo2014} found that the Ni mass depends on $\Delta m_{15}(B)$ and the bolometric peak luminosity. All these
parameters ($m^{max}_{B}$, $\Delta m_{15}(B)$, $t_2$ and $t_L$, and NIR late-phase luminosity) can be tied to the Ni
(and/or Fe) mass in the explosion. We provide a summary of the important correlations analysed in this work for the three NIR filters in Figure \ref{fig:sum}, where we correlate the timing parameters ($t_1$, $t_0$, $t_2$), $
\Delta m_{15}$, $t_L$ and NIR late phase luminosity ($M|_{55}$). With most of the Fe synthesised
in the explosion, we propose that all light curve parameters point to
the Ni as the dominant factor in shaping the light
curves. We find a consistent picture that the  
properties of second maximum in the
NIR light curves are strongly influenced by the amount of Ni produced in
the explosion followed by a more luminous decline.

The NIR colours show a pronounced evolution after t$_1$, the flux in the $J$ band decreases significantly with respect
to both $Y$ and $H$ shortly after t$_1$
(Fig.~\ref{fig:col_scat}). The reason for this depression is most likely the lack of transitions providing the required channel for radiation to
emerge around 1.2 $\mu$m \citep{Spyr1994, Hoeflich1995, Wheeler1998}.
This persists until t$_2$ when the [Fe~II]
(1.25 $\mu$m) emission line forms and the $J$-band magnitude
recovers relative to $H$, although not compared with $Y$, which is also
dominated by Fe lines \citep{Marion2009}. The $H$ filter is
dominated by Co~II lines \citep[e.g.][]{Kasen2006, Jack2012}.  After the
second maximum $J-H$ turns redder again due to the faster decline rate
in the $J$ filter.

We investigated whether the (optically) fast-declining SNe\,Ia can be grouped within our
analysis. These SNe have very low $^{56}$Ni mass which is almost
certainly centrally located \citep[low velocities of Fe~III in the
optical at late times in SN 1991bg;][]{Mazzali1998} and transition so
rapidly that a second maximum barely has time to form. Thus, we could not
include them directly in our study.  However, a comparison of $M|_{55}$
with $\Delta m_{15}(B)$ can be made for four objects of this class (SNe~2005ke, 2006mr, 2007N and 2007ax). The trend of these objects showing a
fainter $M|_{55}$ luminosity with larger $\Delta m_{15}(B)$ is followed,
but it is unclear whether these SN 1991bg-like SNe\,Ia follow the same
relation as their brighter counterparts. The result
remains inconclusive as the scatter remains currently too large.  We
note that the decline rates in the NIR at late times for these objects do
not differ significantly from our sample. They follow the same
distribution as in Fig~\ref{fig:late}.

\subsection{Improved Distance Measurements?}
\label{sec:dist}

The uniformity of the NIR light curves is in use for
cosmological projects \citep{BN12, Weyant2014}. \citet{Krisciunas2009}
finds no correlation of the maximum luminosity with other light curve
parameters, but identified the (optically) fast-declining SNe\,Ia
as subluminous in the NIR compared to the other SNe\,Ia in
their sample. In \citet{Kattner2012}, the authors find a weak trend
between the NIR first maximum luminosity and $\Delta m_{15}$. 

\begin{figure}
\includegraphics[width=.47\textwidth, height=.5\textheight]{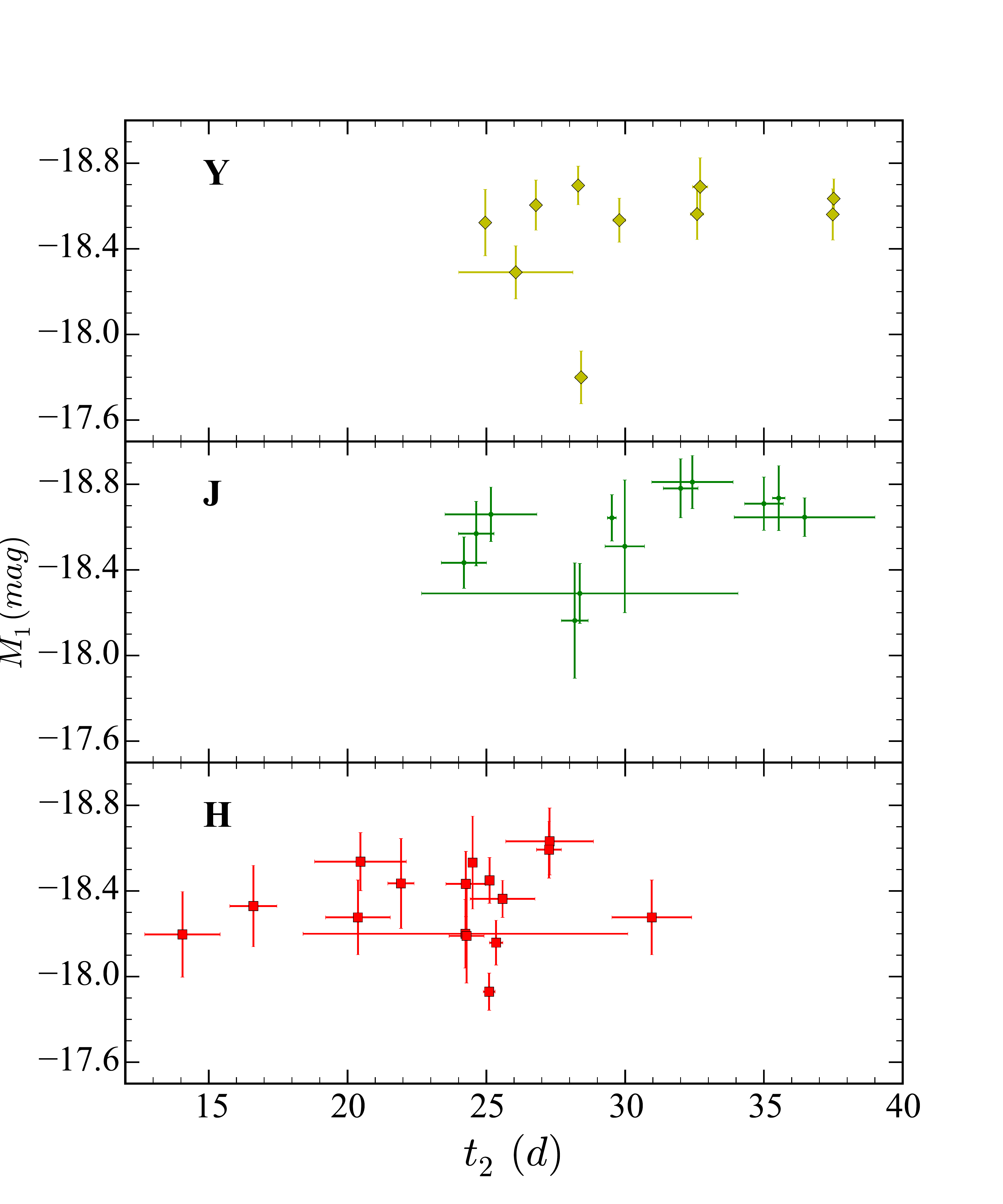}
\caption{
The magnitude of the first maximum in the $YJH$ bands is plotted
against the phase of the second maximum in $J$ band. The underluminous
object in $Y$ and $J$ bands is the heavily extinguished SN\,2006X.  In the $J$ band SN\,2014J also appears fainter.
}
\label{fig:lum}
\end{figure}

We looked
for NIR light curve properties to further improve SNe\,Ia as distance
indicators. In Fig.~\ref{fig:lum} we find no significant evidence for a correlation
between $M_1$ and $t_2$ for objects in our sample. 
The faintness of SN~2006X  in this diagram is probably due to the strong absorption towards this SN.

The scatter in Fig.~\ref{fig:lc1} falls to around 0.2 mag at
later NIR phases. Between t$_0$ and t$_{2}$ all NIR light curves reach comparable luminosities. These phases are between 10 and 20 d in $Y$, near 15 and 20
d in $J$ and $H$ when SNe\,Ia are only about 1 (0.5)
magnitude fainter than at the first maximum in $Y$ ($H$). The decline in
the $J$ light curves is quite steep after the first maximum and the
SNe have already faded by nearly 2 magnitudes. At least for $Y$
and $H$, it might be worthwhile to investigate whether good distances can
be determined at later phases (between +10 and  +15 days). The advantage would be a distance
measurement with a reduced extinction component, i.e. mostly independent
of the exact reddening law, and the possibility of targeted NIR
observations even when the first maximum had been missed.

\section{Conclusions}
\label{sec-conc}
The cosmological interest in SNe~Ia in the NIR stems from the small
observed scatter in the peak magnitudes. We confirm this with our
extended literature sample despite our simple
assumptions on distances and neglect of host galaxy absorption. The phase of the first maximum
in the  $YJH$ light curves shows a narrow distribution.  The
uniformity of the SNe~Ia in the NIR lasts until about one week past
$B$ maximum. The NIR
light curves diverge showing a large scatter by the time of the second
maximum and thereafter. The IR colour curves are uniform at early phases
with increasing scatter after 20 days.

These findings corroborate the use of only few IR observations near the
first maximum to obtain good cosmological distances \citep{BN12,
Weyant2014}. The small scatter prevails for nearly a week around the
first maximum and potentially another window opens near the second
maximum, where the scatter again appears rather small. A condition for
these measurements is accurate phase information to be able to use sparse
NIR observations to derive the distances. The phase
information could come from accurate optical light curves, as used here,
or through spectra. The latter are needed for redshifts and
classification in any case. 

The information contained in the NIR light curves points towards
the nickel/iron mass as the reason for the variations. The absolute
magnitude 55~d past maximum together with the very uniform decline
rate of the light curves in all bands, including the optical
\citep{Barbon1973, Phillips1999, Leibundgut2000}, points towards
differences of the energy input into the ejecta with a rather uniform
density structure. This is also found in theoretical studies
\citep{Kasen2007,Kromer2009,Dessart2014}, although most of these models
employ Chandrasekhar-mass explosions and a wider range of ejecta mass
models may need to be explored to confirm the similarities in the
structure of the ejecta. A higher luminosity at a fixed phase in the
nebular phase points towards a larger iron core and hence a higher
nickel production in the explosion. A corollary is the phase of
the second maximum which also occurs later for larger iron cores
\citep{Kasen2006}. 

The strong
correlation of the phase of the second IR maximum with the optical light
curve shape parameter $\Delta m_{15}$ and the onset of the uniform
$(B-V)$ colour evolution (`Lira law') point towards SNe~Ia as an
ordered family and nickel mass as the dominating factor in shaping the
appearance of SNe~Ia. With a higher nickel mass a larger Fe core is
expected. This would result in higher expansion velocities observed at
late phases \citep{Mazzali1998}. It is worth checking whether this
prediction holds true in future observations.

The phase of the second maximum should provide a handle to determining
the nickel masses in SN~Ia explosions, in particular for SNe~Ia where
absorption is significant. In cases like SN~2014J the NIR light
curve can yield an independent check on the nickel mass and a direct
comparison to the direct measurements of the $\gamma-$rays from the
nickel decay \citep{Churazov2014, Diehl2014}. A calibration of a fair set of
(preferably bolometric) peak luminosities and the derived nickel masses
with our parameter $t_2$ should lead to the corresponding relation.

We attempted to improve the IR first maximum for distance measurements
and looked for a possible correlation of $M_1$ with the phase of the
second maximum. There is
a slight improvement in the $J$ band and none in $Y$ and $H$. The
resulting scatter after correction in $J$ band is 0.16 mag. This is
comparable to the observed value from previous studies
\citep{Mandel2009, Kattner2012, Weyant2014}. 
It remains to be seen whether larger
samples will provide a better handle in the future. 


Since we specifically studied the second maximum in the NIR light
curves, we excluded SNe~Ia, which do not display this feature. These are
objects similar to either SN~1991bg, SN~2000cx or SN~2002cx
and are typically faint and peculiar SNe \citep{Filippenko1992,
Leibundgut1993, Li2001, Li2003, Foley2013}. They appear to also display
a fainter first maximum in the NIR \citep{Krisciunas2009}. It would
be interesting to check whether the uniform decline rate observed in the
near-IR also applies to SNe~Ia which do not show the second maximum. 
This could be used to check whether these explosions share some physical
characteristics with the ones discussed here. There are only very few
SNe~Ia of this type and we do not find a conclusive
answer. Increasing the number of SNe~Ia
with this information is important to assess the physical differences among
the different SN~Ia groups. The recent CFAIR2 catalogue
\citep{Friedman2014} will provide some of these data.

The NIR light curves display a decline rate after the second
maximum, which is significantly larger than the optical light curves at
the same phase \citep[e.g.][]{Leibundgut2000}. At very late phases
($\geq 200$~days) the near-IR light curves become nearly flat as observed
for SN~2001el, while there is no observable change in the optical
decline rates \citep{Stritzinger2007}. It would be interesting to
observe the change in decline rate between 100 and 300 days.
Presumably, the internal structure
of the explosions sets the transition towards the positron
decays as the dominant energy source, when the ejecta have thinned
enough that the $\gamma-$rays escape freely. This phase might be
correlated with other physical parameters, like the nickel and ejecta
mass, determined through early light curves.

A possible extension of this photometric study  with detailed
spectroscopic observations and theoretical spectral synthesis
calculations might be worthwhile to check on the emergence of the various emission lines,
trace the exact transition to the flatter IR light curves and determine
whether it indicates any differences in the structure of the SNe, e.g.
transition to positron channel.

Finally, Euclid will discover many SNe at near-NIR wavelengths out
to cosmologically interesting redshifts \citep[e.g.][]{Astier2014}.  With the small scatter of the
peak luminosity, these observations will provide distances with largely
reduced uncertainties due to reddening. Our study confirms the promise
the NIR observations of SNe~Ia offer.

\section*{Acknowledgements}
\label{sec-ack}

This research was supported by the DFG Cluster of Excellence `Origin and
Structure of the Universe'. We would like to thank Chris Burns for his
help with template fitting using SNooPy, Richard Scalzo for discussion
on the nickel masses and Saraubh Jha on the nature of Type Ia
SNe. We thank St\'ephane Blondin for his comments on the
manuscript.  B.L. acknowledges support for this work by the Deutsche
Forschungsgemeinschaft through the TransRegio project TRR33 `The Dark
Universe' and the Mount Stromlo Observatory for a Distinguished
Visitorship during which most of this publication was prepared.  
S.D. acknowledges the use of University College London computers
Starlink and splinter. K.M. acknowledges support from a Marie Curie 
Intra-European Fellowship, within the 7th European Community Framework Programme (FP7). This research has made use of the NASA/IPAC Extragalactic Database (NED) which is operated by the Jet Propulsion Laboratory, California Institute of Technology, under contract with the National Aeronautics and Space Administration.

\appendix
\section{Parameter Tables for the Correlations}

\begin{table*}
\centering
\caption{$Y$ filter apparent Magnitude and phase of the two maxima and the minimum for the complete sample. The phases are given in days relative to the $B$ maximum. \label{tab:magY}. The full table is available online}
\begin{tabular}{l r r r r r r}
\hline
SN  &   \multicolumn{1}{c}{$t_1$}  & \multicolumn{1}{c}{$m_1$} & \multicolumn{1}{c}{$t_0$}  & \multicolumn{1}{c}{$m_0$} & \multicolumn{1}{c}{$t_2$}  & \multicolumn{1}{c}{$m_2$} \\
    &   \multicolumn{1}{c}{(days)} &                           & \multicolumn{1}{c}{(days)} &                           & \multicolumn{1}{c}{(days)} &                           \\

\hline
2004eo & $-4.48 \pm 0.82$           & $15.90 \pm 0.03$           & \multicolumn{1}{c}{\ldots} & \multicolumn{1}{c}{\ldots} & \multicolumn{1}{c}{\ldots} & \multicolumn{1}{c}{\ldots} \\
2004ey & \multicolumn{1}{c}{\ldots} & \multicolumn{1}{c}{\ldots} & \multicolumn{1}{c}{\ldots} & \multicolumn{1}{c}{\ldots} & $31.77 \pm 0.20$           & $15.51 \pm 0.01$           \\
2004gs & \multicolumn{1}{c}{\ldots} & \multicolumn{1}{c}{\ldots} & $ 8.10 \pm 0.76$           & $17.69 \pm 0.02$           & $22.35 \pm 0.07$           & $17.16 \pm 0.03$           \\
2004gu & \multicolumn{1}{c}{\ldots} & \multicolumn{1}{c}{\ldots} & \multicolumn{1}{c}{\ldots} & \multicolumn{1}{c}{\ldots} & $37.04 \pm 0.13$           & $17.67 \pm 0.03$           \\
2005A  & \multicolumn{1}{c}{\ldots} & \multicolumn{1}{c}{\ldots} & $13.70 \pm 1.61$           & $17.69 \pm 0.41$           & $27.39 \pm 0.07$           & $16.00 \pm 0.11$           \\

\hline
\end{tabular}
\end{table*}

\begin{table*}
\centering
\caption{$J$ filter apparent magnitude and phase of the two maxima and the minimum for the complete sample. The phases are given in days relative to the $B$ maximum. \label{tab:magJ} The full table is available online}
\begin{tabular}{l r r r r r r}
\hline
SN  & \multicolumn{1}{c}{$t_1$}  & \multicolumn{1}{c}{$m_1$} & \multicolumn{1}{c}{$t_0$}  & \multicolumn{1}{c}{$m_0$} & \multicolumn{1}{c}{$t_2$}  & \multicolumn{1}{c}{$m_2$} \\
    & \multicolumn{1}{c}{(days)} &                           & \multicolumn{1}{c}{(days)} &                           & \multicolumn{1}{c}{(days)} &                           \\
\hline
1980N  & \multicolumn{1}{c}{\ldots} & \multicolumn{1}{c}{\ldots} & \multicolumn{1}{c}{\ldots} & \multicolumn{1}{c}{\ldots} & $21.86 \pm 2.70$           & \multicolumn{1}{c}{\ldots} \\
1981B  & \multicolumn{1}{c}{\ldots} & \multicolumn{1}{c}{\ldots} & \multicolumn{1}{c}{\ldots} & \multicolumn{1}{c}{\ldots} & $32.19 \pm 0.10$           & \multicolumn{1}{c}{\ldots} \\
1986G  & \multicolumn{1}{c}{\ldots} & \multicolumn{1}{c}{\ldots} & \multicolumn{1}{c}{\ldots} & \multicolumn{1}{c}{\ldots} & $16.40 \pm 1.40$           & \multicolumn{1}{c}{\ldots} \\
1998bu & \multicolumn{1}{c}{\ldots} & \multicolumn{1}{c}{\ldots} & \multicolumn{1}{c}{\ldots} & \multicolumn{1}{c}{\ldots} & $29.84 \pm 0.40$           & \multicolumn{1}{c}{\ldots} \\
1999ac & \multicolumn{1}{c}{\ldots} & \multicolumn{1}{c}{\ldots} & \multicolumn{1}{c}{\ldots} & \multicolumn{1}{c}{\ldots} & $27.00 \pm 2.00$           & $15.61 \pm 0.50$           \\

\hline
\end{tabular}
\end{table*}
\begin{table*}
\centering
\caption{$H$ filter apparent magnitudes and phases of the two maxima and the minimum for the complete sample. The phases are given in days relative to the $B$ maximum. \label{tab:magH} The full table is available online}
\begin{tabular}{l r r r r r r}
\hline
SN & \multicolumn{1}{c}{$t_1$}   & \multicolumn{1}{c}{$m_1$} & \multicolumn{1}{c}{$t_0$}  & \multicolumn{1}{c}{$m_0$} & \multicolumn{1}{c}{$t_2$}  & \multicolumn{1}{c}{$m_2$} \\
   & \multicolumn{1}{c}{(days)} &                           & \multicolumn{1}{c}{(days)} &                           & \multicolumn{1}{c}{(days)} &                           \\
\hline
1981B  & \multicolumn{1}{c}{\ldots} & \multicolumn{1}{c}{\ldots} & \multicolumn{1}{c}{\ldots} & \multicolumn{1}{c}{\ldots} & $31.98 \pm 0.50$           & \multicolumn{1}{c}{\ldots} \\
1986G  & \multicolumn{1}{c}{\ldots} & \multicolumn{1}{c}{\ldots} & \multicolumn{1}{c}{\ldots} & \multicolumn{1}{c}{\ldots} & $12.70 \pm 1.30$           & $10.01 \pm 0.10$           \\
1998bu & \multicolumn{1}{c}{\ldots} & \multicolumn{1}{c}{\ldots} & \multicolumn{1}{c}{\ldots} & \multicolumn{1}{c}{\ldots} & $29.18 \pm 1.65$           & \multicolumn{1}{c}{\ldots} \\
1999ee & \multicolumn{1}{c}{\ldots} & \multicolumn{1}{c}{\ldots} & \multicolumn{1}{c}{\ldots} & \multicolumn{1}{c}{\ldots} & $30.50 \pm 0.70$           & $15.11 \pm 0.03$           \\
2000E  & \multicolumn{1}{c}{\ldots} & \multicolumn{1}{c}{\ldots} & \multicolumn{1}{c}{\ldots} & \multicolumn{1}{c}{\ldots} & $25.10 \pm 0.70$           & $13.85 \pm 0.10$           \\

\hline
\end{tabular}
\end{table*}

\begin{table*}
\caption{Late time decline rate and luminosity at +55days in $YJH$. The full table is available online}
\begin{tabular}{l c c c c}
\hline
SN & \multicolumn{1}{c}{slope} & \multicolumn{1}{c}{err} &
\multicolumn{1}{c}{$M|_{55}$} & \multicolumn{1}{c}{err} \\
   & \multicolumn{1}{c}{(mag/day)} & & & \\
\hline
Y filter	&		 &		&		&		\\
2005M	&	0.054	&	0.002	& $-17.73$	&	0.08\\
2005el	&	0.056	&	0.002	& $-16.72$	&	0.11\\
2005na	&	0.052	&	0.003	& $-17.57$	&	0.13\\
2006D	&	0.050	&	0.001	& $-16.62$	&	0.17\\
2006X	&	0.054	&	0.003	& $-16.67$       &       0.10\\ 
\hline
\end{tabular}
\label{tab:late_decl}
\end{table*}
\begin{table*}
\caption{$t_2$ and epoch at which SN enters Lira law. The full table is available online}
\begin{tabular}{l c c c c}
\hline
SN & $t_2$ & err & $t_{L}$ & err \\
   & (days) &     & (days) &     \\
\hline
Y filter	&		&	 	&		&	 \\
2004gs	&	22.34	&	0.07	&	24.17	&	3.21	\\ 
2005am	&	21.73	&	0.12	&	22.69	&	1.26	\\ 
2005el	&	24.96	&	0.11	&	25.78	&	1.00	\\ 
2005na	&	31.77	&	3.46	&	32.69	&	1.26	\\ 
2006D	&	24.88	&	0.02	&	25.00	&	0.26	\\ 
\hline
\end{tabular}
\label{tab:lira}
\end{table*}


\label{lastpage}
\end{document}